\documentclass{mn2e}			

\usepackage{amsfonts}
\usepackage{amsmath}
\usepackage{graphicx}

\title[Curvature in scaling relations]
      {Curvature in the scaling relations of early-type galaxies}

\author[J. Hyde \& M. Bernardi]
{Joseph B. Hyde \& Mariangela Bernardi\thanks{E-mail: jhyde,bernardm@physics.upenn.edu}\\
      Department of Physics \& Astronomy, University of Pennsylvania, 
      209 S. 33rd St., Philadelphia, PA 19104, USA}

\newcommand{\aj}[1]{AJ}
\newcommand{\apj}[1]{ApJ}
\newcommand{\apjl}[1]{ApJL}
\newcommand{\apjs}[1]{ApJS}
\newcommand{\mnras}[1]{MNRAS}

\begin{document}
\pagerange{\pageref{firstpage}--\pageref{lastpage}}

\maketitle

\label{firstpage}

\begin{abstract}
We select a sample of about 50,000 early-type galaxies from the 
Sloan Digital Sky Survey (SDSS), 
calibrate fitting formulae which correct for known problems with 
photometric reductions of extended objects, apply these corrections, 
and then measure a number of pairwise scaling relations in the 
corrected sample.  
We show that, because they are not seeing corrected, the use of 
Petrosian-based quantities in magnitude limited surveys leads to 
biases, and suggest that this is one reason why Petrosian-based 
analyses of BCGs have failed to find significant differences from 
the bulk of the early-type population.  
These biases are not present when seeing-corrected parameters 
derived from deVaucouleur fits are used.  
Most of the scaling relations we study show evidence for curvature:  
the most luminous galaxies have smaller velocity dispersions, 
larger sizes, and fainter surface brightnesses than expected if 
there were no curvature.  
These statements remain true if we replace luminosities with 
stellar masses; they suggest that dissipation is less important at 
the massive end.  
There is curvature in the dynamical to stellar mass relation as 
well:  the ratio of dynamical to stellar mass increases as stellar 
mass increases, but it curves upwards from this scaling both at 
small and large stellar masses.  
In all cases, the curvature at low masses becomes apparent when 
the sample becomes dominated by objects with stellar masses smaller 
than $3\times 10^{10}M_\odot$.   
We quantify all these trends using second order polynomials; these 
generally provide significantly better description of the data than 
linear fits, except at the least luminous end.
\end{abstract}


\section{Introduction}
Early-type galaxy observables, luminosities $L$, 
half-light radii $R_e$, mean surface brightnesses $I_e$, 
colors, and velocity dispersions $\sigma$, 
are strongly correlated with one another.  
These scaling relations are usually described as single 
power-laws ($R_e\propto L^{3/5}$, $\sigma\propto L^{1/4}$, 
            $g-r\propto \sigma^{0.4}$, $R_e\propto I_e^{-0.8}$), 
suggesting a single formation mechanism across the population.  
However, galaxy formation models suggest that the brightest galaxies 
in clusters had unusual formation histories (De Lucia et al. 2006; 
Almeida et al. 2008), so they may follow different scaling laws 
(Boylan-Kolchin et al. 2005; Robertson et al. 2006;
Hopkins et al. 2008a).  
Formation histories, and the importance of gaseous dissipation 
and/or gas rich mergers are also expected to have been different 
depending on the mass range of the galaxy 
(e.g. Mihos \& Hernquist 1993; Naab et al. 2006; Hopkins et al. 2008b, Ciotti et al. 2007).
So, one might reasonably expect to see departures from the single 
power-law scaling relations, especially at the extremes of the 
population.  

Measurements of brightest cluster galaxy scaling relations have 
shown them to be unusual (Malumuth \& Kirshner 1981, 1985;
Oegerle \& Hoessel 1991; Postman \& Lauer 1995).  
And statistically significant detections of curvature in many 
scaling relations across the entire population have now been made 
(Zaritsky et al. 2006; Lauer et al. 2007; Bernardi et al. 2007; 
Desroches et al. 2007; Liu et al. 2008; but see 
von der Linden et al. 2007).  
The main goal of this paper is to exploit the large sample size 
provided by the Sloan Digital Sky Survey (hereafter SDSS) to 
make precision measurements of the curvature in these scaling 
relations.

Section~\ref{sample} describes how we select a sample of early-type 
galaxies from the SDSS.  It also discusses the corrections we apply 
to account for the fact that the SDSS photometric reductions are 
unreliable for extended objects.  
These are particularly important since the curvature we would 
like to measure is small (else it would have been seen in smaller 
samples), so the photometric and spectroscopic parameters at the 
extremes of the sample must be reliable.  

Section~\ref{curved} quantifies the curvature in a number of 
pairwise scaling relations for this sample.  It also shows that, 
in the relations which involve luminosity, the curvature persists 
if one replaces luminosities with stellar masses.  
A final section summarizes our findings.  

An Appendix discusses why, because they are not seeing-corrected, 
Petrosian based quantities are ill-suited for precision measurements 
in deep, magnitude-limited, ground-based datasets.  It also shows why 
the Petrosian-based analysis of BCGs by von der Linden et al. (2007) 
yielded anomalous results.  

\section{The Sample}\label{sample}
We start from a sample of 376471 galaxies based on the Fourth Data 
Release (DR4) of the SDSS but with parameters updated to the SDSS DR6 
values (Adelman-McCarthy et al. 2008).
From this sample we extract 46410 early-type galaxies following the 
procedure described below. 
We use SDSS deVaucouleur magnitudes and sizes, {\tt model} colors, and 
aperture corrected velocity dispersions to $r_e/8$ unless stated otherwise.  
The SDSS also outputs Petrosian magnitudes and sizes.  
However, these are not seeing corrected, and Appendix~A shows that 
this introduces systematic biases, so we do not use them in what 
follows.
Throughout, angular diameter and luminosity distances were 
computed from the measured redshifts assuming a Hubble constant 
of $H_0=70$~km~s$^{-1}$~Mpc$^{-1}$ 
in a geometrically flat background model dominated by a 
cosmological constant at the present time:
  $(\Omega_0,\Lambda_0) = (0.3,0.7)$. 

About notation: we use $R_e$ to specify radius in kpc, and $r_e$ to specify
angular size in arcseconds. For surface brightnesses, we use the following definition.
\begin{eqnarray}
\mu_e &=& -2.5\log_{10}(I_e) = -2.5\log_{10} \left(\frac{L}{2\pi R_e^2}\right) \nonumber\\ 
&=& m + 5\log_{10}(r_e)+2.5\log_{10}(2\pi) - 10\log_{10}(1+z)
\end{eqnarray}
where $m$ is the evolution, reddening, and k-corrected apparent magnitude. We use $M_x$ 
to denote the absolute magnitude in band $x$, and $M_*$ to denote stellar mass in solar units.

\begin{figure}
 \centering
 \includegraphics[width=\hsize]{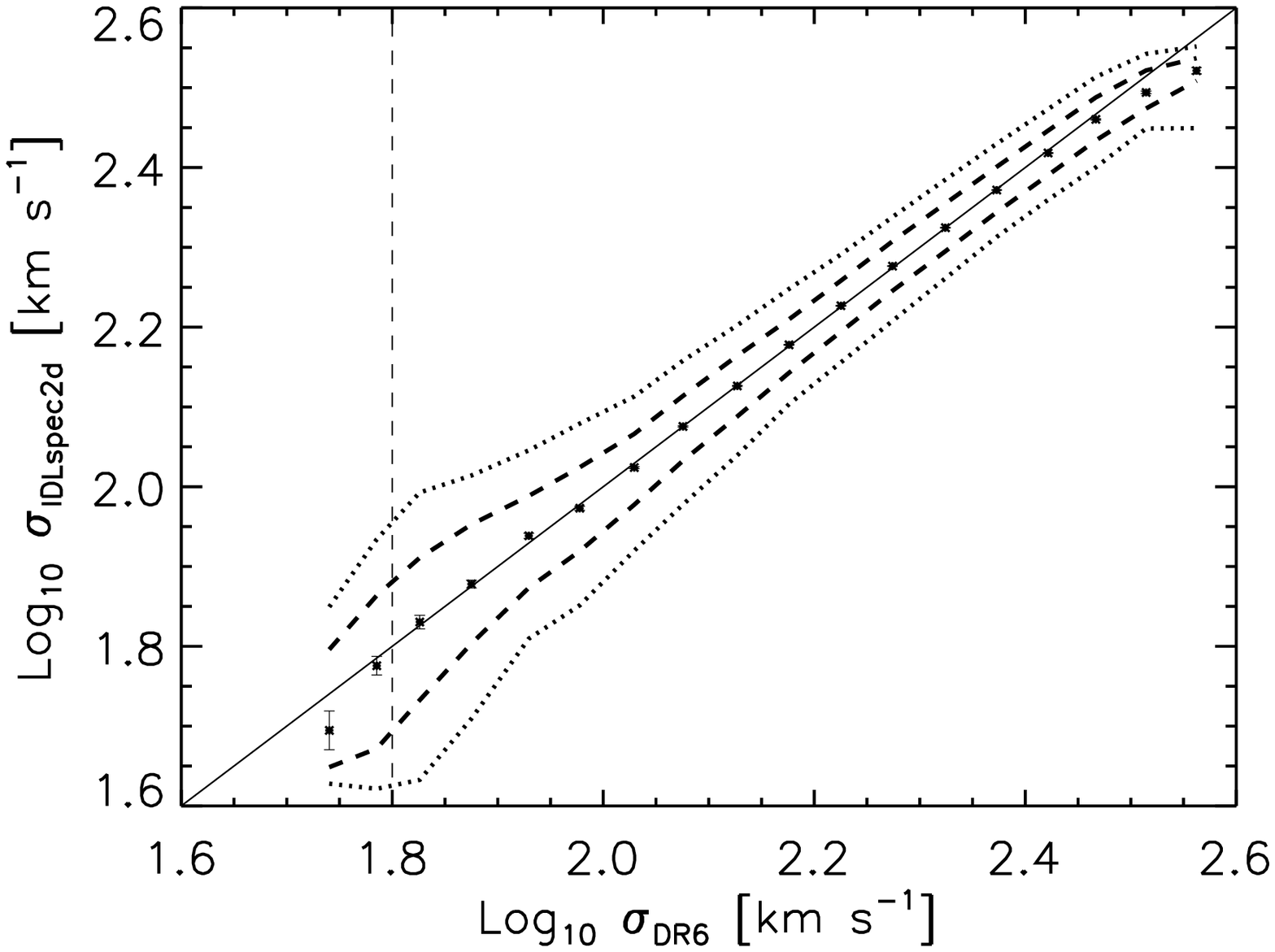}
 \includegraphics[width=\hsize]{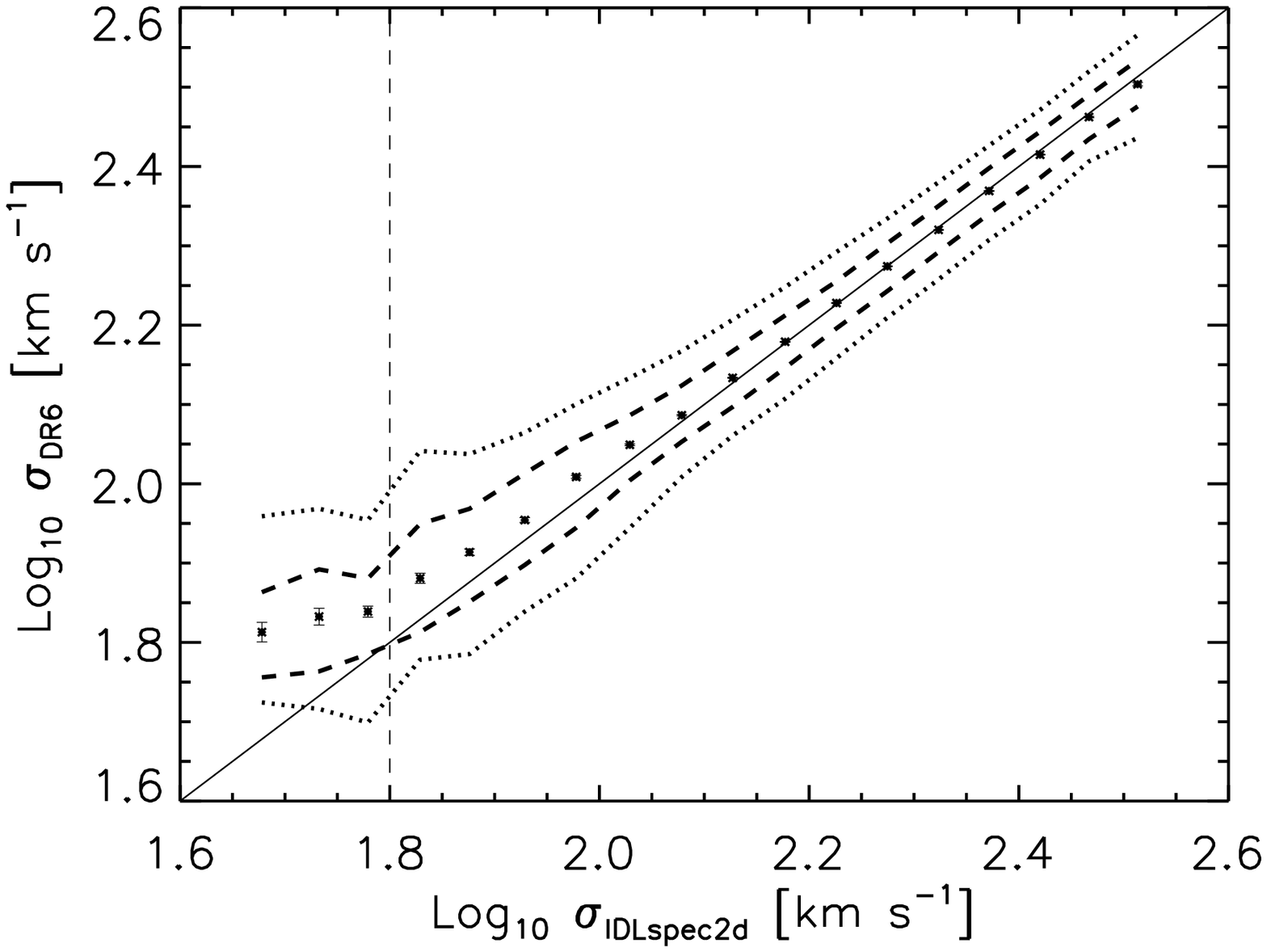}
 \caption{Comparison of the velocity dispersion measurements from the
SDSS-DR6 and the {\tt IDLspec2d} reduction. Vertical dashed line
shows our cut at small $\sigma\sim 60$~km s$^{-1}$. Symbols with 
error bars show the median value and its uncertainty, 
dashed and dotted curves show the regions which enclose 68\% and 95\% 
of the objects.
          }
 \label{vdisp}
\end{figure}

\subsection{Selecting early-types}
To obtain a sample of early-type galaxies we first select the 
subset of galaxies which are very well-fit by a deVaucouleur 
profile in the $g$ and $r$ bands ($g$-band {\tt fracDev} = 1 
and $r$-band {\tt fracDev} = 1); this gives about 100603 objects. 
To avoid contamination by later-type galaxies we also require
the spectrum to be of ``early-type'', by setting eClass $< 0$ 
(see SDSS documentation for details of this classification).
This slightly reduces the sample, to 94934 galaxies.
Since the SDSS spectroscopic survey is magnitude limited, we 
require that $r-$band deVaucouleur magnitudes satisfy
 $14.5 < m_r < 17.5$. (Spectra are actually taken for objects 
having Petrosian magnitude $m_{r,Pet}<17.77$; our more conservative 
cut is designed to account for the fact that the Petrosian quantity 
systematically underestimates the total light in a deVaucouleur 
profile.)  This cut leaves 70417 galaxies.  Of these, 64492 have 
stellar mass estimates from Gallazzi et al. (2005).

We would also like to study the velocity dispersions of these objects.
One of the important differences between the SDSS-DR6 and previous 
releases is that the low velocity dispersions ($\sigma< 150$~km~s$^{-1}$) 
were biased high; this has been corrected in the DR6 release 
(see DR6 documentation, or discussion in Bernardi 2007).
We compared the values given by the official SDSS-DR6 database and those 
computed in the Princeton reduction ({\tt IDLspec2d}). The two sets of 
measurements still show weak systematic trends. The upper panel of 
Figure~\ref{vdisp} shows that at fixed $\sigma_{\rm DR6}$ the median 
values of $\sigma_{\rm IDLspec2d}$ agree quite well with the 
$\sigma_{\rm DR6}$ values, except at large 
$\sigma_{\rm DR6} > 320$ km s$^{-1}$, where $\sigma_{\rm DR6}$ is 
slightly larger.  However, the bottom panel shows that, when compared 
at fixed $\sigma_{\rm IDLspec2d}$, a systematic trend is more 
evident -- especially at small $\sigma_{\rm IDLspec2d} < 120$ km s$^{-1}$.
To minimize systematics, we decided to average the DR6 and IDLspec2d 
velocity dispersion measurements.  

\begin{figure*}
 \centering
 \includegraphics[width=0.49\hsize]{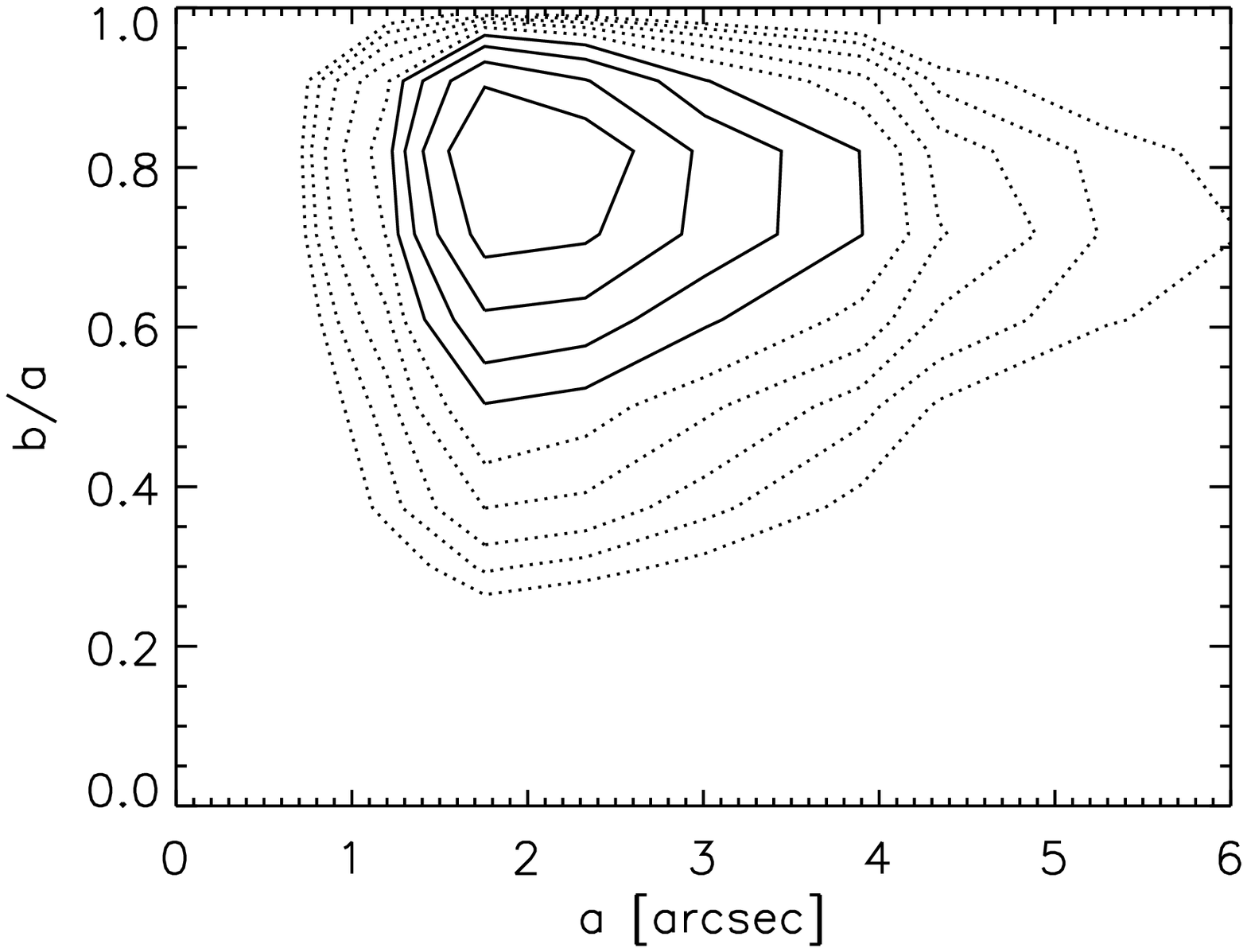}
 \includegraphics[width=0.49\hsize]{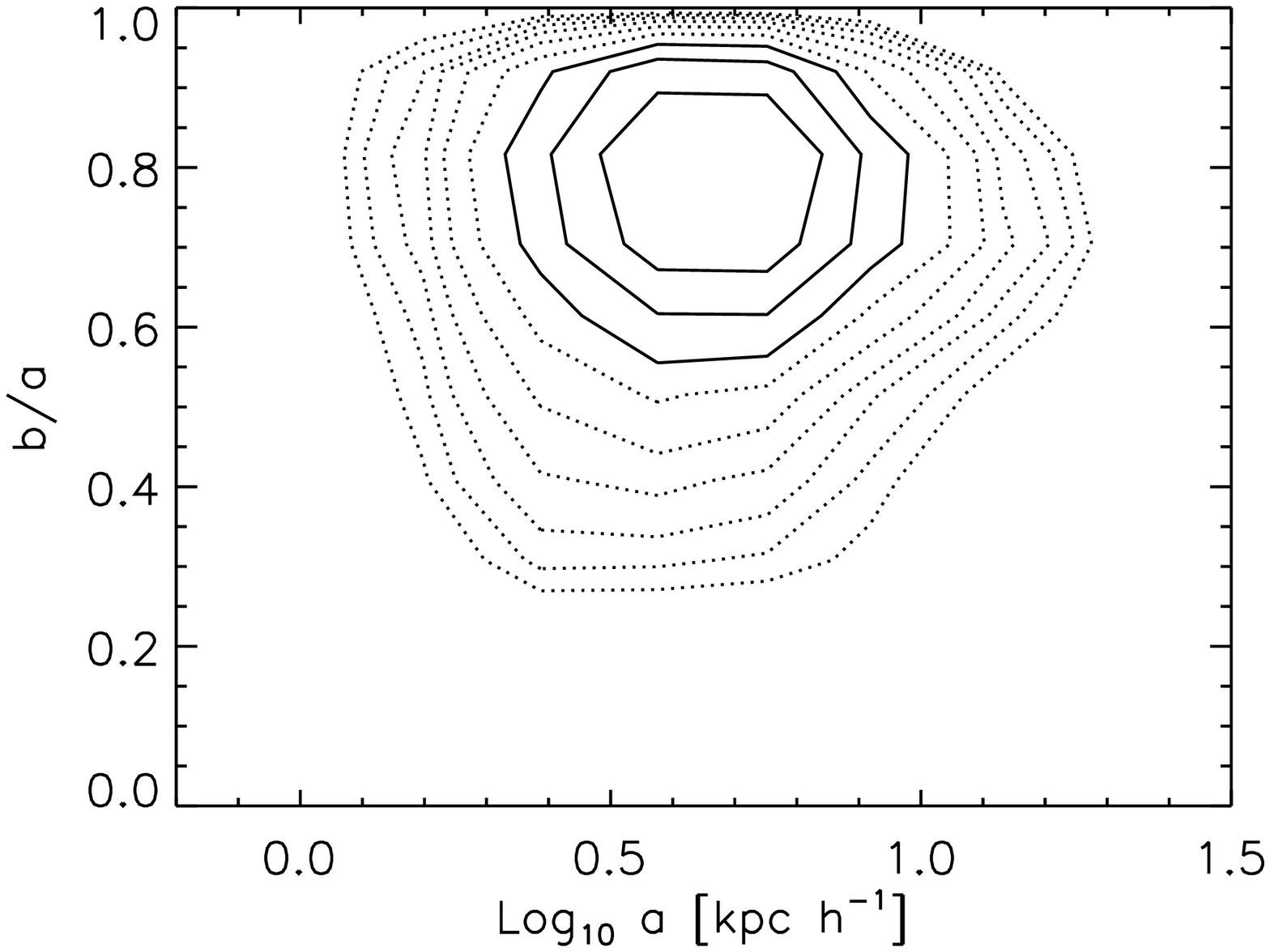}
 \caption{Distribution of $b/a$ in the $r$-band as a function of 
          angular (left) and physical size (right).  Contours 
          show regions of equal probability density. Starting at maximum density, 
		each line represents a factor of $\sqrt{2}$ decrease in density.  The change from solid to dotted 
          lines marks the point which encloses 68\% of the sample. 
          There appears to be a separate population of small 
          $b/a < 0.6$ objects at physical sizes smaller than 
          $a = 10h^{-1}$~kpc.}
 \label{ba}
\end{figure*}

At the low end, we select galaxies with $\sigma > 60$~km~s$^{-1}$
(see Section~\ref{contam}, Bernardi et al. 2003c and Hyde \& Bernardi 2008 
for discussion of biases introduced by eliminating objects based on 
their $\sigma$).  
At the high end we select galaxies with $\sigma < 400$~km s$^{-1}$ 
to avoid contamination from double/multiple superpositions (see 
Bernardi et al. 2006b, 2008).  The maximum $\sigma$ of a single galaxy 
given in Bernardi et al. 2008, and confirmed by Salviander et al. 2008 
using high resolution spectra, is about 430~km~s$^{-1}$.
In addition, the SDSS-DR6 only reports velocity dispersions if the $S/N$ 
in the spectrum in the restframe $4000-5700$~\AA\ is larger than 10 or with
the {\tt status} flag equal to 4 (i.e. this tends to exclude galaxies with
emission lines). 
To avoid introducing a bias from these cuts, we have also estimated 
velocity dispersions for all the remaining objects.  It turns out that 
this cut does not change the correlations studied here (nor those in 
the Fundamental Plane study of Hyde \& Bernardi 2008).
The net result is to change our sample from 64492 to 64343 objects.
About 25 objects have colors which lie outside the range 
$0.4\le g-r\le 1.3$ which we do not believe, so we exclude them 
from the sample.  

\begin{figure}
 \centering
 \includegraphics[width=0.95\hsize]{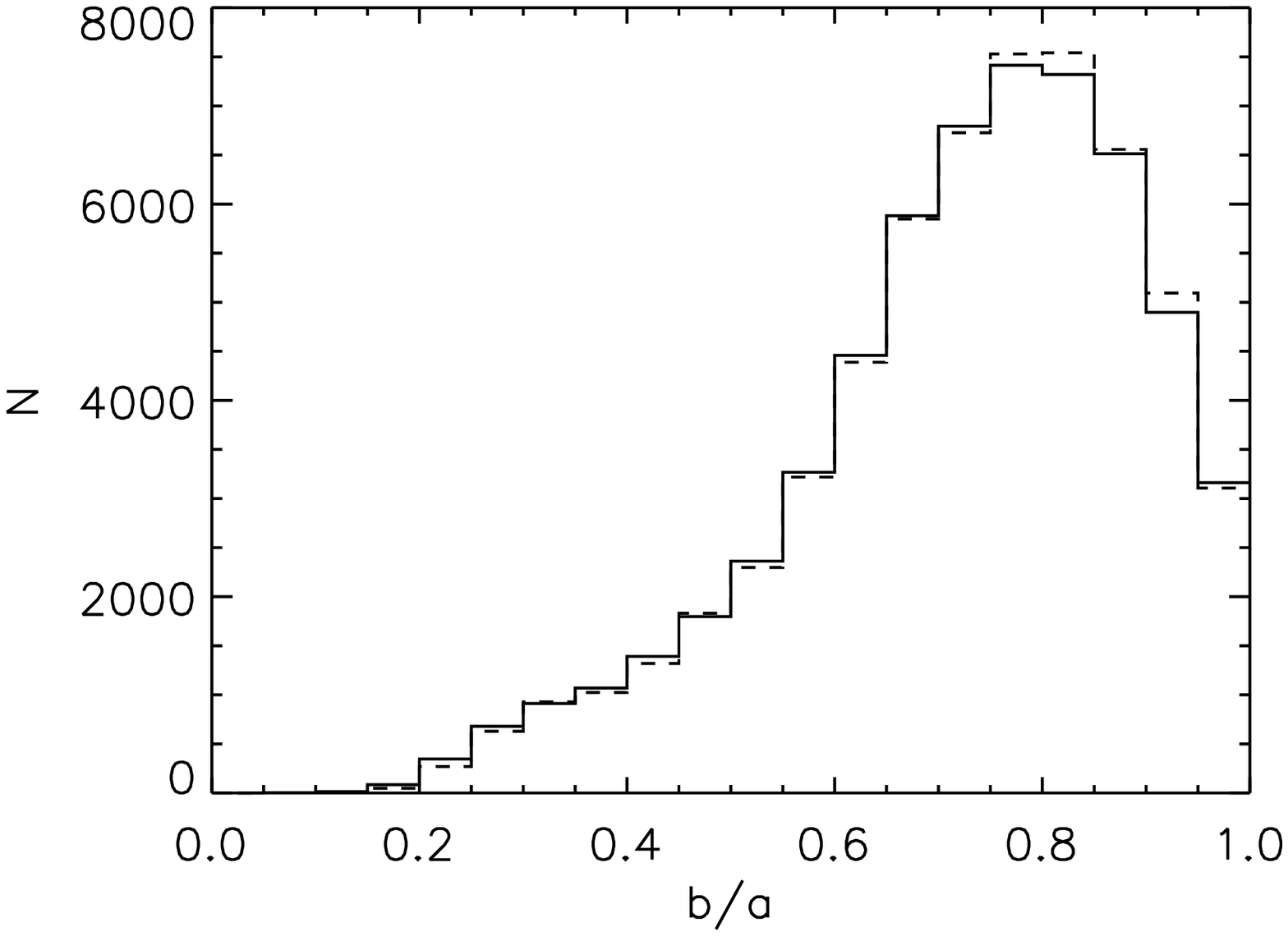}
 \caption{Histogram showing distribution of $b/a$ in the 
          $g-$ (solid) and $r-$ (dashed) bands.}
 \label{histba}
\end{figure}

\begin{figure}
 \centering
 \includegraphics[width=\hsize]{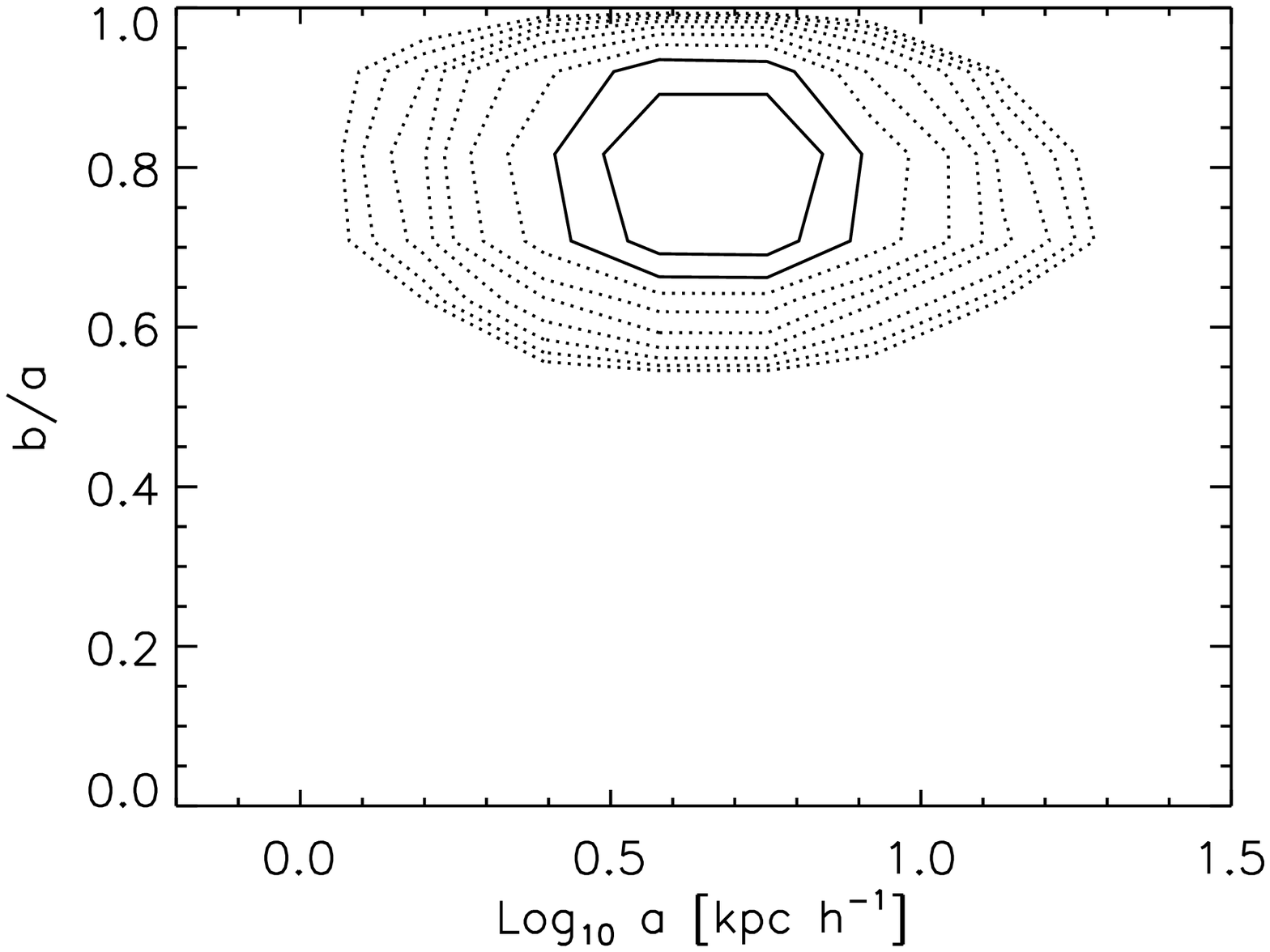}
 \caption{Dependence of axis-ratio $b/a$ on $a$ in the $r-$band
          for galaxies with $b/a > 0.6$ in the $g-$band.  
          Contours show the joint distribution of $b/a$ and $a$ 
          with the same conventions for spacing 
          and line styles as in Figure~\ref{ba}.  
          }
 \label{bar}
\end{figure}

Figure~\ref{ba} suggests that we should make one more cut.  
The left hand panel shows the distribution of axis ratios $b/a$ 
in the sample, as a function of the angular half light radius (left) 
and physical scale (right).  There is clear evidence for two 
populations, particularly when plotted as a function of physical 
scale:  one with $b/a > 0.6$ and the other with $b/a < 0.6$.  
The objects with small $b/a$ make up about twenty percent of 
our full early-type sample, but they are a larger fraction of 
the smaller fainter galaxies than they are of the bright.  
There are good physical reasons to suspect that these objects are 
a different population (e.g. rotational support is necessary if 
$b/a<0.6$), so, in what follows, we remove all objects with 
$b/a<0.6$ in the $g-$band from the sample.  This leaves 51379
objects.  Many of the figures which follow show quantities in the 
$r-$band, for which this cut appears less sharp.
Figure~\ref{histba} shows that the distribution of $b/a$ in the 
two bands is very similar: the small differences between $b/a$ in 
$g$ and $r$ is due, in part, to measurement error.  

Figure~\ref{bar} shows $b/a$ vs $a$ in the $r-$band 
in our sample after applying this cut.  
Comparison with Figure~\ref{ba} shows that the `second' population 
at low $b/a$ has been removed.  In the sample which remains, 
there is a weak tendency for the largest galaxies to have 
slightly smaller $b/a$.  The implications are discussed further 
in Bernardi et al. (2008) who show that the mean $b/a$ drops 
slightly at the highest luminosities and $\sigma$. 

Finally, we must account for the fact that objects were brighter 
in the past by $0.9\,z$~mags because of stellar evolution (e.g. Bernardi et al. 2003a).
This, with $k$-corrections, adjusts slightly the actual values of the magnitude limit which 
we should use when computing effective volumes.  
In fact, these effects work in opposite directions, so the net 
effect is small.  However, since our goal is to quantify small 
effects in a large sample, it is necessary to do this carefully. 
The net result is to reduce the sample size by about ten percent, 
to 46410.  As a check that this has been done correctly, we perform 
the test suggested by Schmidt (1968).  
If $V_i$ is the survey volume between object $i$ and the observer, 
and $V_{{\rm max},i}$ is the total survey volume over which the 
object could have been seen, then the average value of 
$(V/V_{\rm max})$ should be 0.5:  
we find $\langle V/V_{\rm max}\rangle = 0.499$.  
(The luminosity evolution is slightly larger than, but statistically 
consistent with, the $0.85z$~mags reported by Bernardi et al. 2003b.)  

These cuts are similar to those described by Bernardi et al. (2003a, 2006a), 
who provide further details about the motivation for each cut,  
except for: i) the cut on $b/a$; ii) the inclusion of velocity dispersions 
from low $S/N$ spectra or with the {\tt status} flag not-equal to 4; and iii) the inclusion of 
velocity dispersions at $60 < \sigma < 90$~km~s$^{-1}$.
In addition, for the present study, we have chosen to be more 
conservative.  Previously, we required Petrosian magnitudes 
$14.5 < m_{r,{\rm Pet}} < 17.75$.  Changing to a brighter limit 
makes our final sample size considerably smaller than if we had 
used the Bernardi et al. (2006a) selection.   
In addition, we previously required {\tt fracDev} $> 0.8$ in the 
$r-$band; we now require {\tt fracDev} $= 1$ in $r$ as well as in $g$, 
because non-early-type features are expected to be more obvious in 
the $g-$ band.  
This more conservative choice for {\tt fracDev} eliminated about 
20000 objects (doing the selection based on $g$ but not $r$ makes 
little difference, because requiring {\tt fracDev} $= 1$ is quite 
stringent).  To quantify the effect of these additional cuts, we 
have applied them to the Bernardi et al. (2006a) sample, and found 
that the comoving number density of objects is reduced to about 0.4 
times that in Bernardi et al. (2006a).  
None of the results which follow are sensitive to the value of the 
comoving number density.  

\begin{figure}
 \centering
 \includegraphics[width=\hsize]{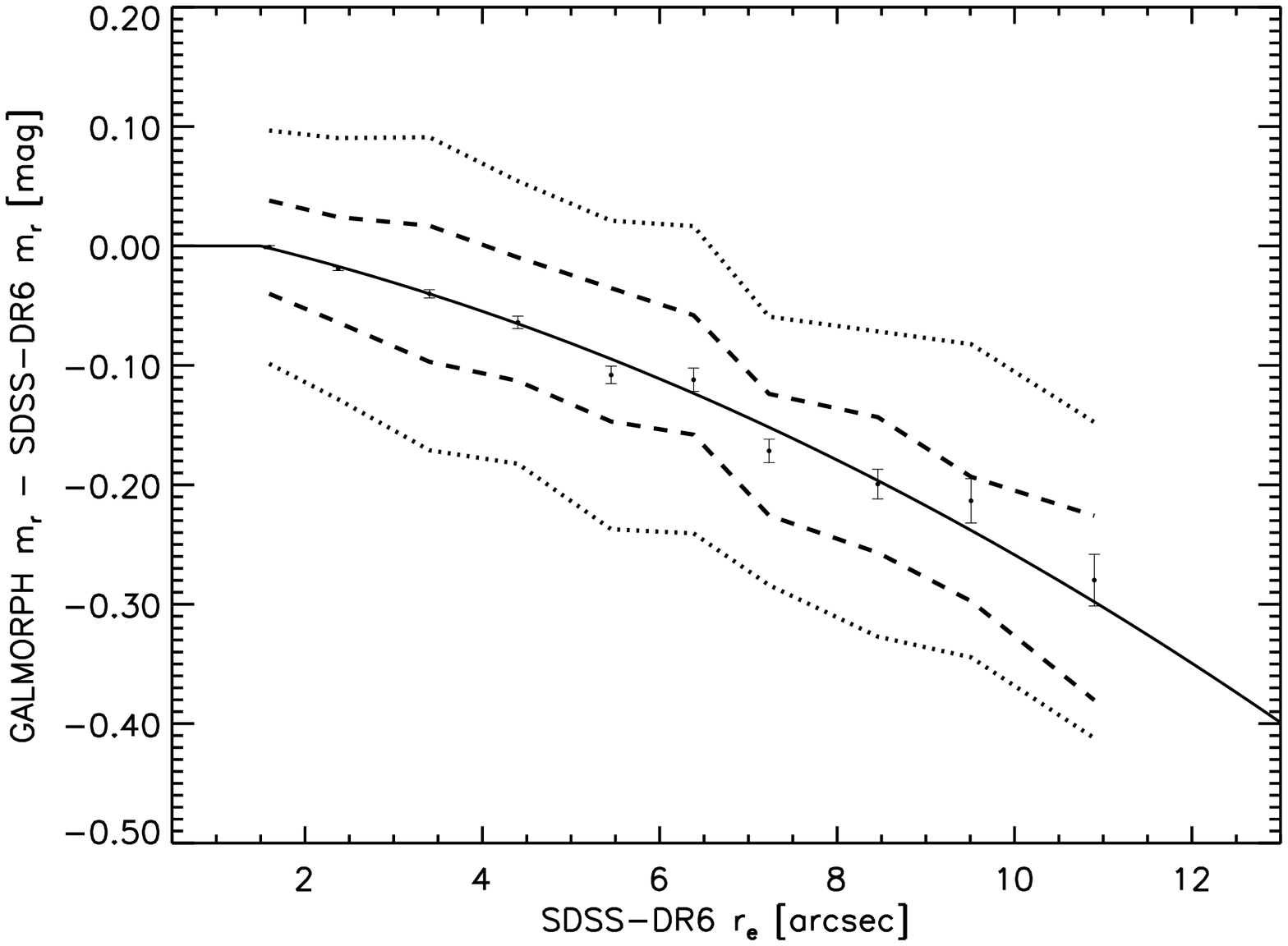}
 \includegraphics[width=\hsize]{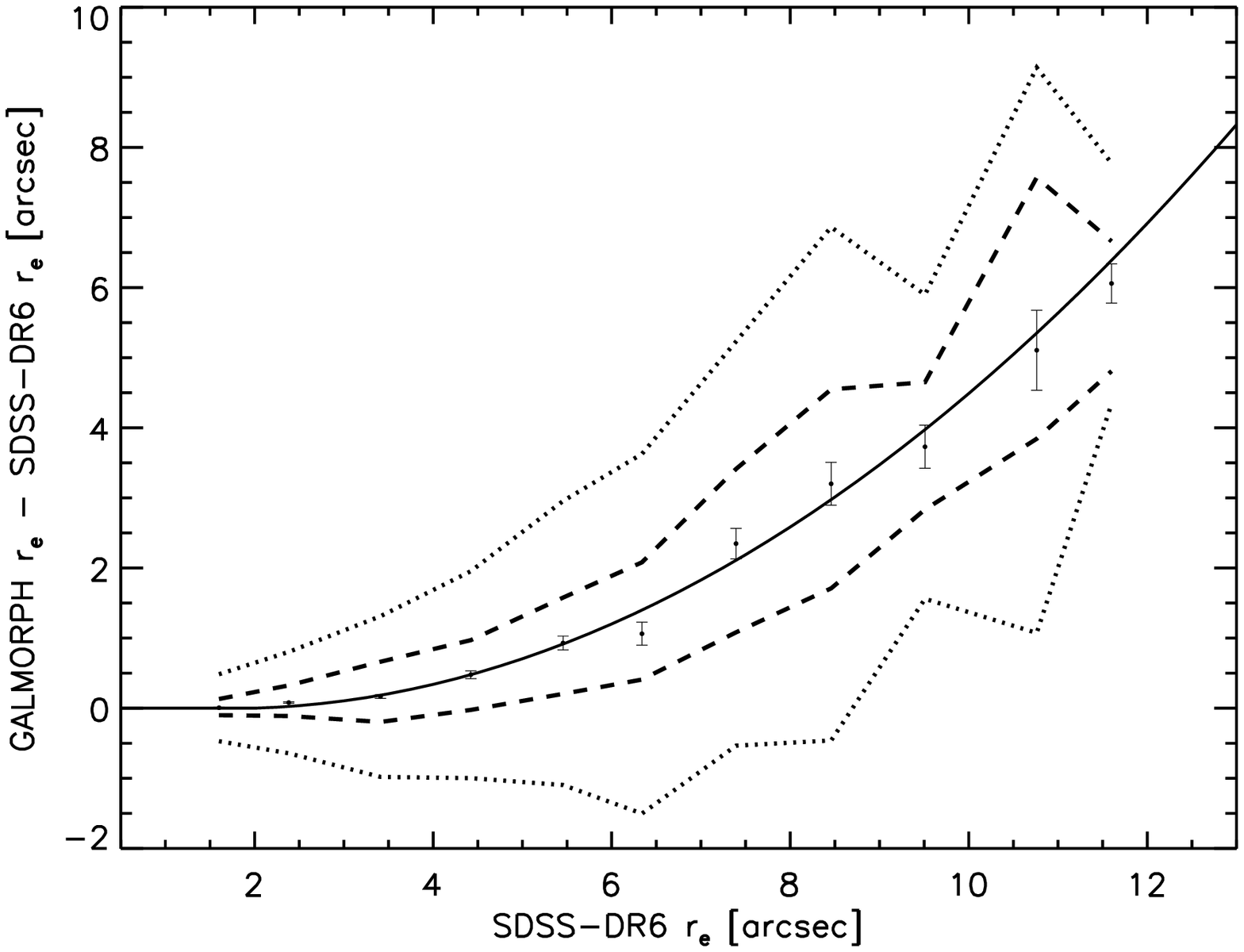}
 \includegraphics[width=\hsize]{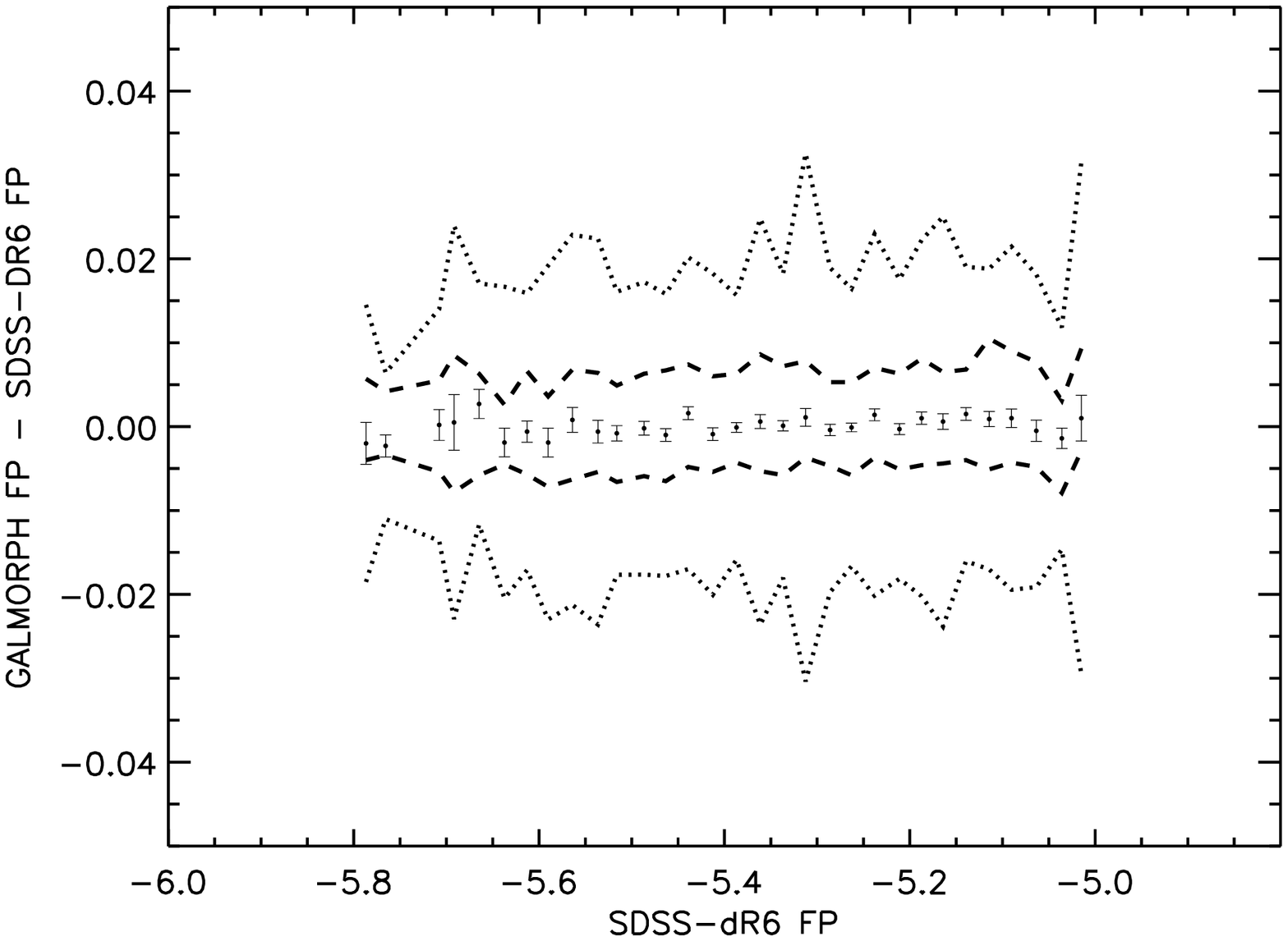}
 \caption{Comparison of GALMORPH and SDSS photometric reductions.  
          Although the two are in good agreement at small $r_e$, 
          where the bulk of the objects lie, the SDSS underestimates 
          the total flux (top) and the half-light radius (middle) 
          of large objects. 
          However, both pipelines return consistent values of 
          the quantity $\log_{10} ( r_e ) - 0.3\,\mu_e$ 
          which Saglia et al. (2001) call FP.  
          Symbols with error bars show the mean relation and its 
          uncertainty, dashed and dotted curves show the regions 
          which enclose 68\% and 95\% of the objects, and smooth 
          solid curves show the fits given by 
          equations~(\ref{mfit}) and~(\ref{rfit}).
         }
 \label{gmorph-sdss}
\end{figure}

\begin{figure}
 \centering
 \includegraphics[width=\hsize]{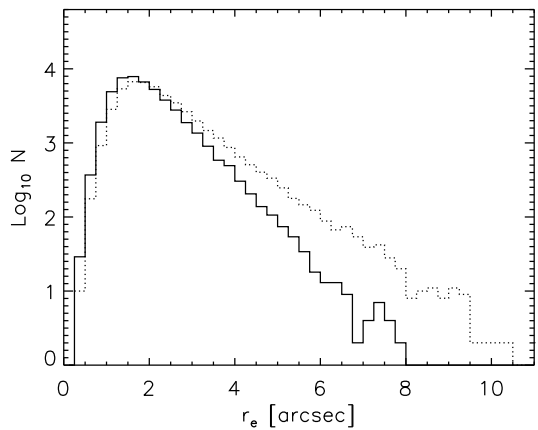}
 \caption{Distribution of sizes obtained by the SDSS 
          photometric reduction (solid) and corrected 
          following Eq.~\ref{rfit} (dotted).
          Although the two are in good agreement at small $r_e$, 
          where the bulk of the objects lie, the corrected sizes 
          are larger for the small fraction of large $r_e$.}
 \label{histR}
\end{figure}

\subsection{Corrections to photometry}
We must address another issue before proceeding.  This is because 
the SDSS reductions are known to suffer from sky subtraction 
problems, particularly for large objects (Adelman-McCarthy et al. 2008).  To illustrate, 
Figure~\ref{gmorph-sdss} compares SDSS photometric reductions 
with those of our own code, GALMORPH, for a 
subset of the full sample ($\sim 5500$ galaxies used by 
Bernardi et al. 2003a plus $\sim 180$ Bright Cluster Galaxies 
analyzed by Bernardi et al. 2007). The GALMORPH reductions do not 
suffer from the sky subtraction problem. This figure shows that 
while the two pipelines are in excellent agreement for small 
objects, the SDSS underestimates the sizes and brightnesses of 
large objects. However, the quantity IP 
$=\log{10} ( r_e ) - 0.3\,\mu_e$, identified by Saglia et al. (2001)
(they called it FP), is not substantially changed.
In all the panels, symbols with error bars show the mean relation 
and the error on the mean, dashed curves show the region 
encompassing 68\% of the objects, and dotted curves enclose 95\%.  

Unfortunately, it is computationally expensive to run GALMORPH 
on the entire DR6 data release.  Therefore, we have fit smooth 
curves to the trends shown in Figure~\ref{gmorph-sdss} (solid 
curves) and we use these fits to correct the SDSS reductions as 
follows.  Given $r_{\rm SDSS}$ we set 
\begin{eqnarray}
 r_e &=& r_{\rm SDSS} + \Delta r_{\rm fit}(r_{\rm SDSS}) \\
 \label{rnew}
 m &=& m_{\rm SDSS} + \Delta m_{\rm fit}(r_{\rm SDSS}),
 \label{mnew}
\end{eqnarray}

where $r_{\rm SDSS}=r_{deV} \sqrt{ \frac{b}{a} }$ is the SDSS-measured deVaucouleur
radius expressed as a geometric mean ($\sqrt{ab}$) of the semimajor ($a$) and semiminor ($b$) axes of 
a half-light containing ellipse. $r_{deV}$, the semimajor axis length, and $\frac{b}{a}$, the axis ratio, are obtained from 
the SDSS catalogs where they are referred to as ``deVRAD'' and ``devAB'', respectively.
$\Delta r_{\rm fit}(r_{\rm SDSS})$ is zero if $r_{\rm SDSS} < 2$~arcseconds and $\Delta m_{\rm fit}(r_{\rm SDSS})$
is zero if $r_{\rm SDSS} < 1.5$~arcseconds, otherwise
\begin{eqnarray}
 \frac{\Delta m_{\rm fit}(r_{\rm SDSS})}{\rm mags} 
  &=& 0.024279 - \frac{r_{\rm SDSS}}{71.1734} 
               - \left(\frac{r_{\rm SDSS}}{26.5}\right)^2
 \label{mfit}\\
 \frac{\Delta r_{\rm fit}(r_{\rm SDSS})}{\rm arcsec} 
  &=& 0.181571 - \frac{r_{\rm SDSS}}{4.5213} 
               + \left(\frac{r_{\rm SDSS}}{3.9165}\right)^2
 \label{rfit}
\end{eqnarray}
We propagate the errors similarly.
Note that $-0.6\,\ln(10)\,r_{\rm SDSS}\,\Delta m_{\rm fit}$ 
provides a good approximation to $\Delta r_{\rm fit}$, as 
one might expect given the bottom panel in Figure~\ref{gmorph-sdss};
it slightly underestimates $\Delta r_{\rm fit}$ at large $r_{\rm SDSS}$. 
We have tested these corrections, and found them applicable to the SDSS $g$, $r$, and $i$ bands.  
Throughout the paper we will denote angular size in arcseconds with $r_e$ 
and physical size in ${\rm h^{-1}\ kpc}$ with $R_e$. These values refer to geometric-mean
deVaucouleur radii, corrected as described in this section and Section~\ref{s_size_corr}.

Later in this paper, we will use stellar mass estimates from 
Gallazzi et al. (2005).  These are derived by estimating a 
stellar mass-to-light ratio, and then multiplying by the estimated 
luminosity.  Since our corrections to the apparent magnitudes will 
affect these luminosities (they increase slightly on average), 
we add equation~(3) to Gallazzi et al. stellar mass estimates (accounting
for the conversion from luminosity to magnitude, $m = -2.5 \log_{10} L$).  
The net effect is to slightly increase some of the stellar masses, 
but we have not included a plot showing this increase.

Figure~\ref{histR} shows the distribution of the effective radii 
of the 46410 early-type galaxies (i.e., whatever their $b/a$ value) 
before and after applying the correction given in equation~(\ref{rfit}).
Although the two are in good agreement at small $r_e$, where the 
bulk of the objects lie, the corrected sizes are larger for the 
small fraction of objects with large $r_e$.

\begin{figure}
 \centering
 \includegraphics[width=0.95\hsize]{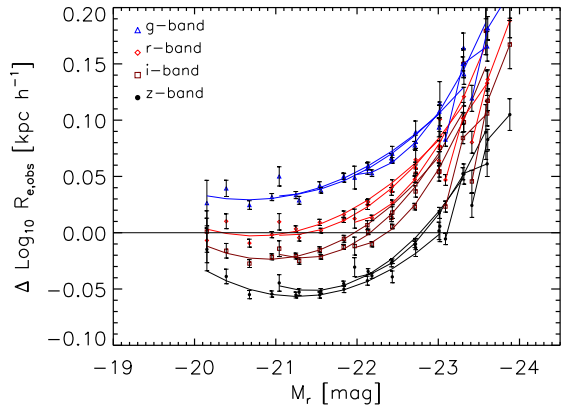}
 \includegraphics[width=0.95\hsize]{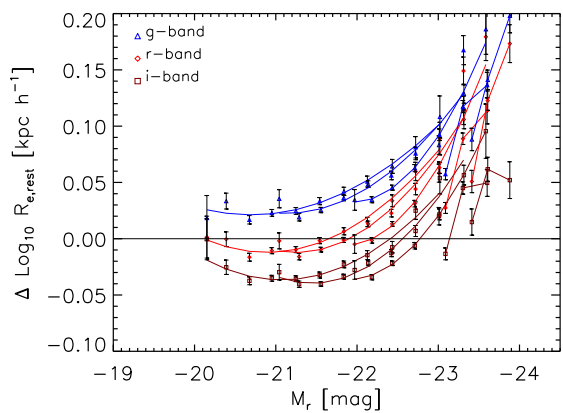}
 \caption{Observed (top) and corrected (bottom) sizes in the SDSS
          derived from deVaucouleur fits to the light profiles shown as a function 
          of $r$-band luminosity $M_r$.  (In all cases,
          $\Delta \log_{10} R_e \equiv \log_{10}R_e + 0.22\,M_r + 4.24$.)
	      $g$-, $r$-, $i$-, and $z$-bands are shown from top to bottom
		  in the upper panel. The bottom panel omits the $z$-band.
		  In each panel, for each band, different lines show data from 
          the following redshift bins:  
          $0.07< z\le 0.1$, $0.1< z\le 0.13$, $0.15< z\le 0.18$, 
          $0.22< z\le 0.25$ and $0.25< z\le 0.35$.  
          }
\label{rkcorr}
\end{figure}

\subsection{Corrections to sizes}
\label{s_size_corr}
We make one final correction to the sizes, which is aimed at 
accounting for the fact that the early-type galaxies have 
color gradients:  on average, their optical half-light radii 
are larger in bluer bands.  If not accounted for, a population 
of intrinsically identical objects will appear to be slightly 
but systematically larger if they are at higher redshift.  
Our sample is large enough that we must correct for this 
effect, the moral equivalent of the $k$-correction to the 
luminosities.  We do so following Bernardi et al. (2003a).

\begin{table*}
 \caption[]{Coefficients of best-fitting relations of the form 
            $\langle Y|X\rangle = p_0 + p_1\, X + p_2\, X^2$ to 
            pairwise scaling relations.  Fits were made to the 
            binned points, not to the objects themselves.  
            Linear fits ($p_2=0$) were made to the galaxies and
            restricted to the range
            $10.5 < \log_{10}(M_*/M_\odot) < 11.5$ and 
            $-23< M_r < -20.5$. The errors on the fitted coefficients are 
            random errors:  they depend on slope of the relation, its scatter, 
            and the sample size.  These are smaller than those produced by 
            systematic effects (e.g., using $\sigma_{\rm DR6}$ 
            or $\sigma_{\rm IDLspec2d}$ rather than their average). 
            Typical systematics errors are a few times larger than
            the random errors.
}
 \centering
 \begin{tabular}{crrrr}
  \hline && \\
  Relation & $p_0\ \ $ & $p_1\ \ $ & $p_2\ \ $ & $\Delta \chi^2_\nu $\\
  \hline && \\
$\langle R|M_*\rangle$ & $-4.79\pm0.02$ & $0.489\pm0.002$ & $-\ \ $ &  $-\ \ $ \\
$\langle R|M_*\rangle$ & $7.55\pm0.44$ & $-1.84\pm0.08$ & $0.110\pm0.004$ & $44.39$ \\
$\langle V|M_*\rangle$ & $-0.86\pm0.02$ & $0.286\pm0.002$ &   $-\ \ $ &  $-\ \ $ \\
$\langle V|M_*\rangle$ & $-5.97\pm0.27$ & $1.24\pm0.05$ & $-0.044\pm0.002$ & $19.20$\\
$\langle I_*|M_*\rangle$ & $-21.77\pm0.09$  & $-0.077\pm0.009$ & $-\ \ $  &  $-\ \ $  \\
$\langle I_*|M_*\rangle$ & $42.11\pm2.23$ & $-12.13\pm0.41$ & $0.57\pm0.02$ & $48.37$ \\
$\langle R|M\rangle$ & $-4.24\pm0.02$ & $-0.221\pm0.001$ & $-\ \ $ &  $-\ \ $  \\
$\langle R|M\rangle$ & $4.72\pm0.32$ & $0.63\pm0.03$ & $0.020\pm0.001$ & $38.70$\\
$\langle V|M\rangle$ & $-0.32\pm0.01$ & $-0.119\pm0.001$ & $-\ \ $ &  $-\ \ $  \\
$\langle V|M\rangle$ & $-2.97\pm0.23$ & $-0.37\pm0.02$ & $-0.006\pm0.001$ & $6.21$\\
$\langle I|M\rangle$ & $17.37\pm0.08$ & $-0.104\pm0.004$ & $-\ \ $ &  $-\ \ $ \\
$\langle I|M\rangle$ & $61.57\pm1.61$ & $4.10\pm0.15$ &  $0.099\pm0.003$ & $37.96$\\
\hline && \\
  $\langle M_{dyn}/L_r|L_r\rangle$ & $-1.50\pm0.03 $ & $0.200\pm0.003 $ & $-\ \ $ &  $-\ \ $ \\
  $\langle M_{dyn}/L_r|L_r\rangle$ & $5.34\pm0.70 $ & $-1.10\pm0.13 $ & $0.062\pm0.001 $ & $5.74$ \\
  $\langle M_{dyn}/M_*|M_*\rangle$ & $-0.46\pm0.02 $ & $0.062\pm0.002 $ & $-\ \ $ &  $-\ \ $ \\
  $\langle M_{dyn}/M_*|M_*\rangle$ & $2.25\pm0.55 $ & $-0.48\pm0.10 $ & $0.027\pm0.005 $ & $1.44$ \\
\hline && \\
$\langle R|V\rangle$ & $-1.42\pm0.02$ & $0.835\pm0.008$ &  $-\ \ $ &  $-\ \ $ \\
$\langle R|V\rangle$ & $2.46\pm0.23$ & $-2.79\pm0.20$ & $0.84\pm0.05$ & $12.16$\\
$\langle I|R\rangle$ & $-4.41\pm0.03$ & $0.246\pm0.001$ &  $-\ \ $ &  $-\ \ $ \\
$\langle I|R\rangle$ & $-24.60\pm0.57$& $2.30\pm0.06$ & $-0.052\pm0.002$ & $62.77$\\
\hline && \\
\end{tabular}
\label{tab:parabolas} 
\end{table*}

We estimate the rest-frame radius in each band by interpolating
the observed radii in adjacent bands. For example, to estimate the rest-frame
$g$-band radius, we use the following expression:
\begin{equation}
R_{e,g,{\rm rest}} = \frac{(1+z) \lambda_g - \lambda_r} {\lambda_g - \lambda_r}\left(R_{e,g,{\rm obs}} - R_{e,r,{\rm obs}}\right)  + R_{e,r,{\rm obs}}
\end{equation}
where $\lambda_{g,r,i,z} = \{4686{\rm \AA},6165{\rm \AA},7481{\rm \AA},8931{\rm \AA}\}$ are the average wavelengths of the SDSS filters,
and $z$ is the spectroscopically-determined redshift.

Figure~\ref{rkcorr} shows why this correction is necessary.  
It shows the sizes of objects in different bands as a function of 
$r$-band ($k$- and evolution-corrected) luminosity.  To reduce the 
range of sizes, we have subtracted-off a luminosity dependent factor:  
we actually show
 $\Delta \log_{10}R_e \equiv \log_{10}(R_e/{\rm kpc}) + 0.221\,M_r + 4.239$ 
(the reason for the exact choice of parameters will become 
clear shortly -- see Table~1 -- but note that, for the present 
purpose, the exact choice is not important).  
The upper panel shows the observed sizes:  
from top to bottom, the sets of symbols and lines show $g$-, $r$-, $i$-, and $z$-bands.
For each band, different lines show data from a number of redshift bins:  
$0.07< z\le 0.1$, $0.1< z\le 0.13$, $0.15< z\le 0.18$, 
$0.22< z\le 0.25$ and $0.25< z\le 0.35$.  
In any redshift bin, the sizes are clearly larger in $g$- and 
smaller in $i$- than they are in the $r$-band.  
The bottom panel shows rest-frame sizes, for $g$-, $r$-, and $i$-bands.
We omit the $z$-band since longer wavelength observations would be necessary
to reconstruct the $z$-band rest-frame size.
The rest-frame sizes are indeed larger in the 
bluer bands, with the difference perhaps slightly smaller at large 
luminosity.  The redshift dependence of the size-luminosity relation is
not apparent when using the the observed radii (upper-panel), but can be seen
when using the rest-frame radii (lower panel); this is studied further in Bernardi (2009).  

The SDSS also outputs non-parametric Petrosian sizes.  
However, in contrast to the deVaucouleur-fitted quantities, 
these sizes are not corrected for the effects of seeing.  
The Appendix shows that, in contrast to the deVaucouleur sizes, 
the Petrosian sizes of objects at higher redshift are systematically 
larger than those from the model based fits.  This is not surprising 
if seeing has compromised the Petrosian-based measurements, suggesting 
that, if this is not accounted for, then the use of Petrosian sizes 
limits or biases the precision measurements which large sample sizes 
would otherwise allow.  

\section{Curvature in scaling relations}\label{curved}
We now turn to measurements of a number of scaling relations.  
It turns out that curvature is often even more obvious if we 
replace luminosity with stellar mass, so we will often show 
such relations side by side.  Unless we specify otherwise, the 
luminosity is always from the $r$-band. We will sometimes
use a shorthand for the $r$-band quantities: $R,V,I,M$ for 
$\log_{10} R_e $, $\log_{10} \sigma $, $\mu_e$, $M_r$.

\subsection{Curvature in pairwise relations}

\begin{figure*}
 \vspace{-2cm}
 \centering
 \includegraphics[width=0.475\hsize]{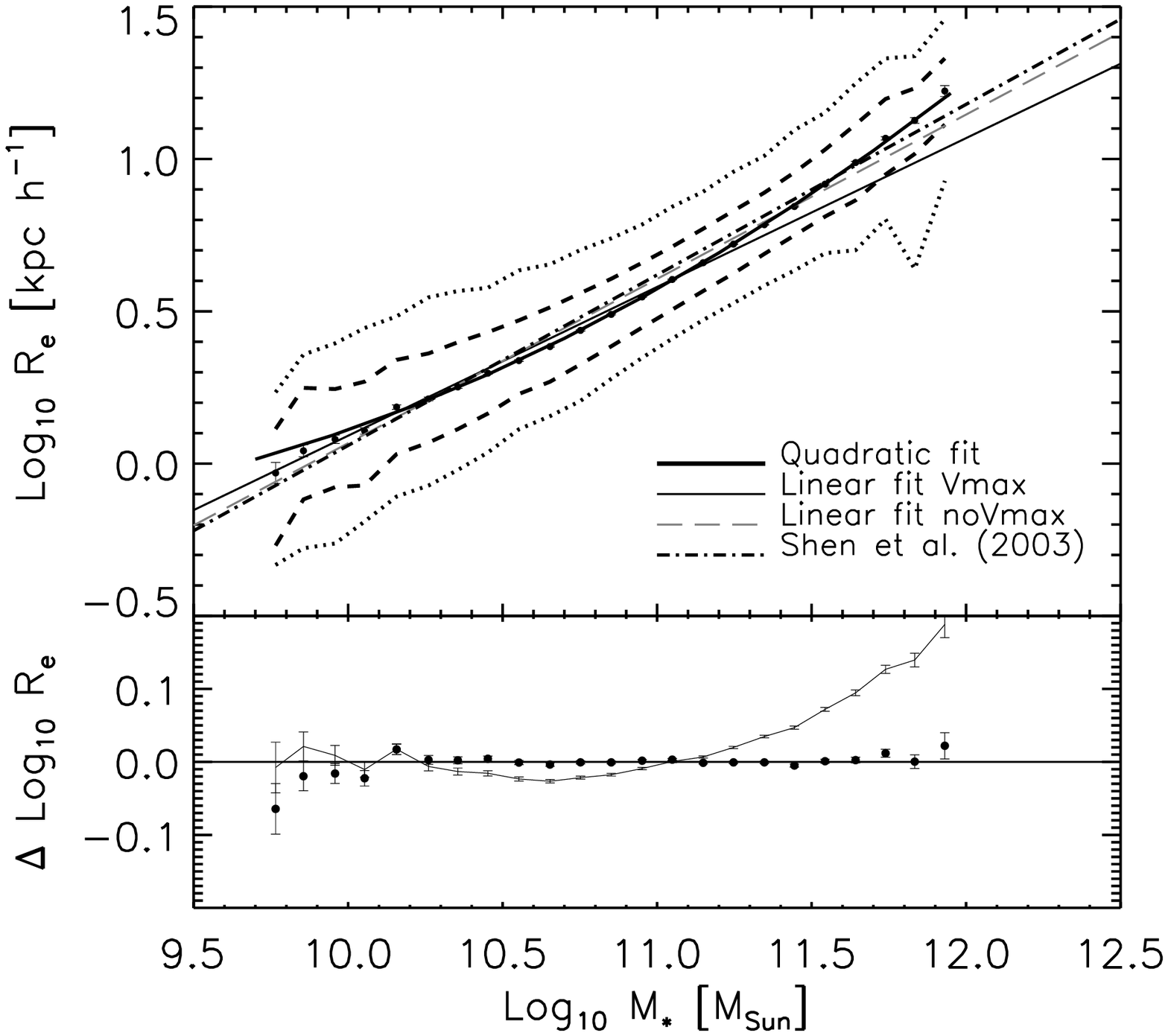}
 \includegraphics[width=0.475\hsize]{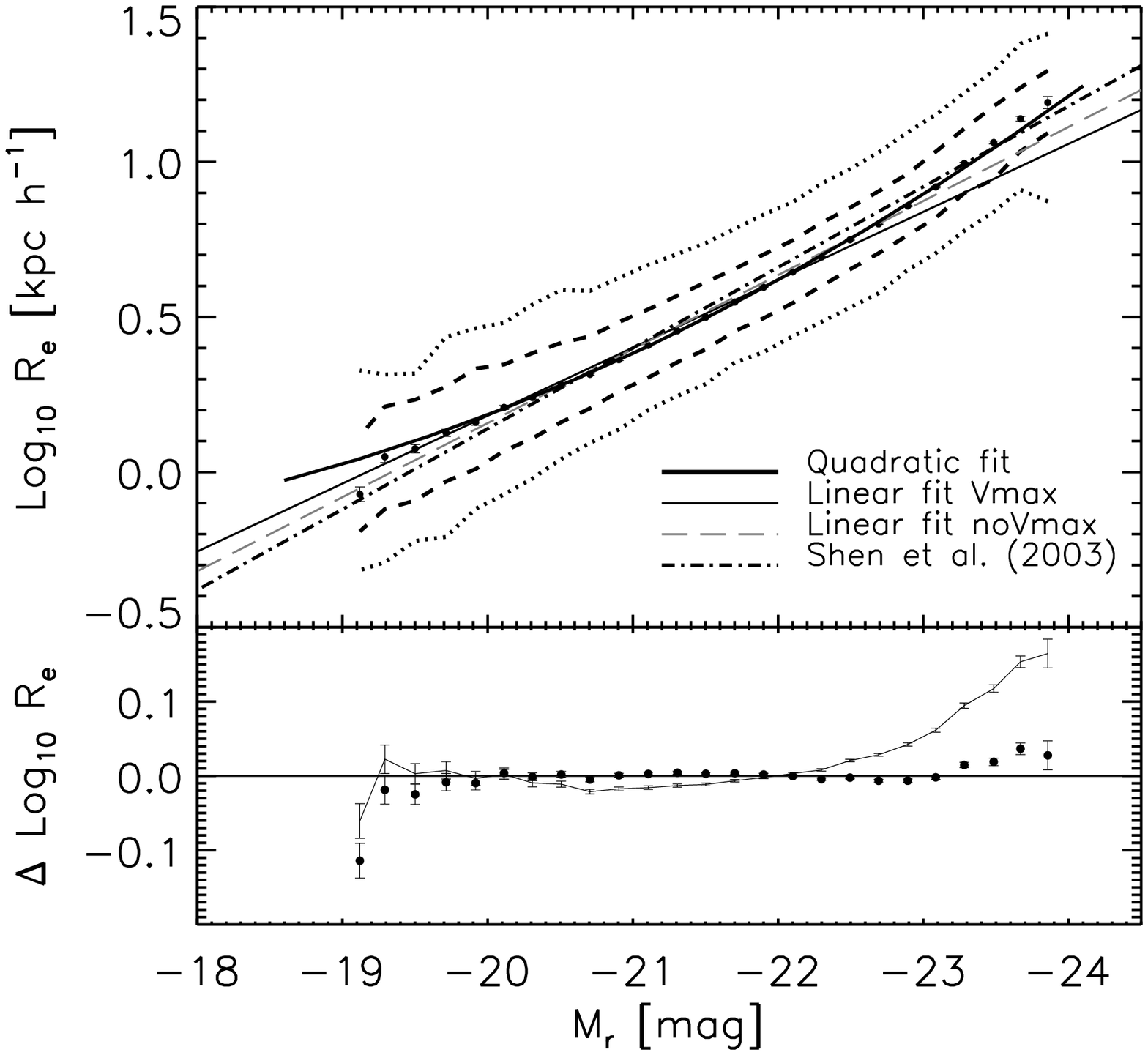}
 \caption{Half-light radius vs stellar mass and luminosity 
          (left and right). 
          Symbols with error bars show the median in small  
          mass or luminosity bins, and dashed and dotted lines 
          show the regions which contain 68\% and 95\% of the 
          objects in each bin.  
          Curves show fits of the form 
            $\langle Y|X\rangle = p_0 + p_1\, X + p_2\, X^2$ to 
          these relations; best-fit coefficients are provided in 
          Table~\ref{tab:parabolas}.  Curved fits were made to the binned 
          counts (symbols with error bars), rather than to the 
          objects themselves.  
          Straight lines show linear fits to these relations; 
          these were made to the galaxies and restricted to the range
          $10.5 < \log_{10}(M_*/M_\odot) < 11.5$ and 
          $-23< M_r < -20.5$.  
          Bottom panels show residuals from the 
          $V_{\rm max}^{-1}$-weighted linear (connected thin line) and 
          quadratic fits (filled circles).}
 \label{RL_f}
\end{figure*}

\begin{figure*}
 \vspace{-2cm}
 \centering
 \includegraphics[width=0.475\hsize]{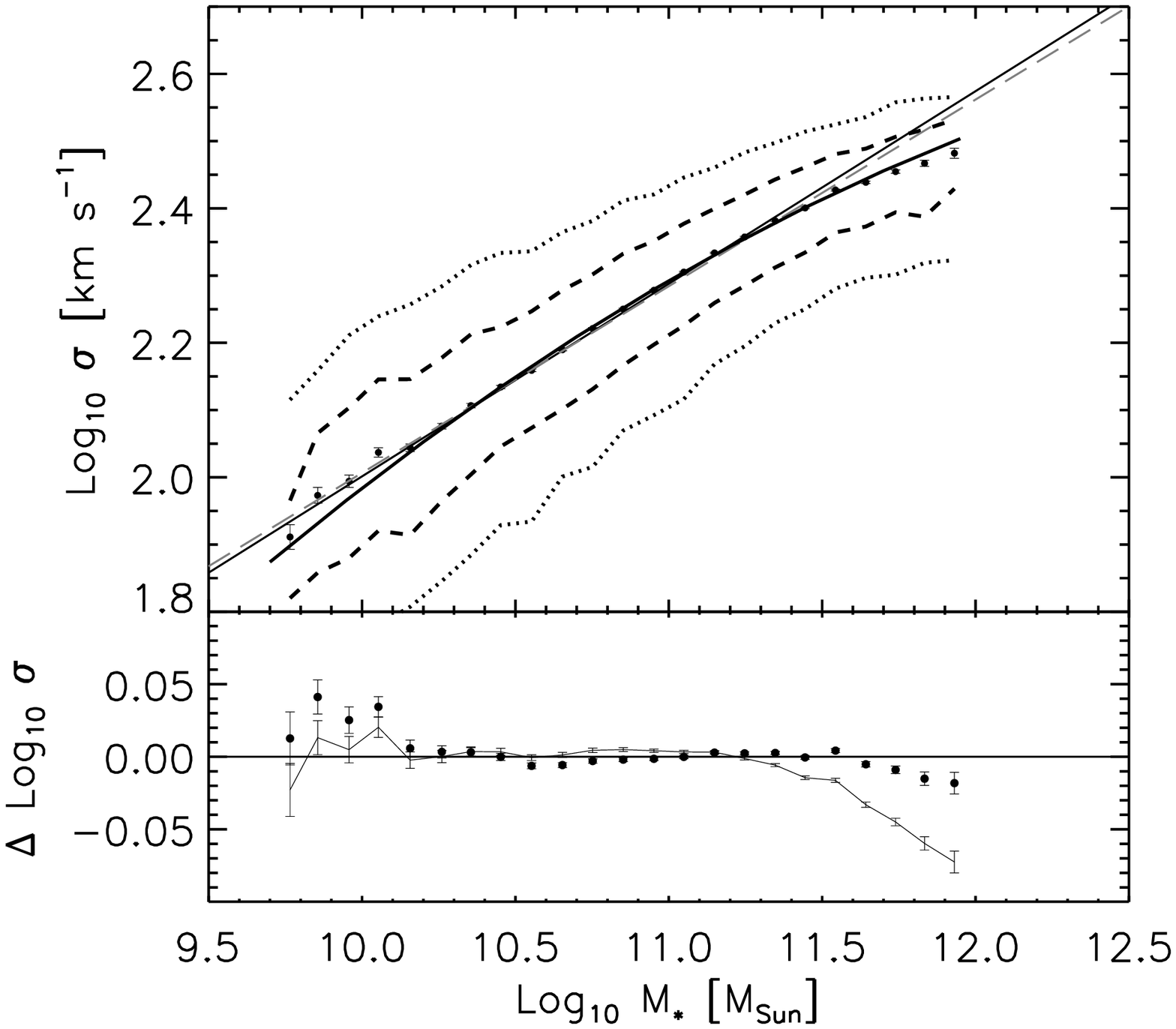}
 \includegraphics[width=0.475\hsize]{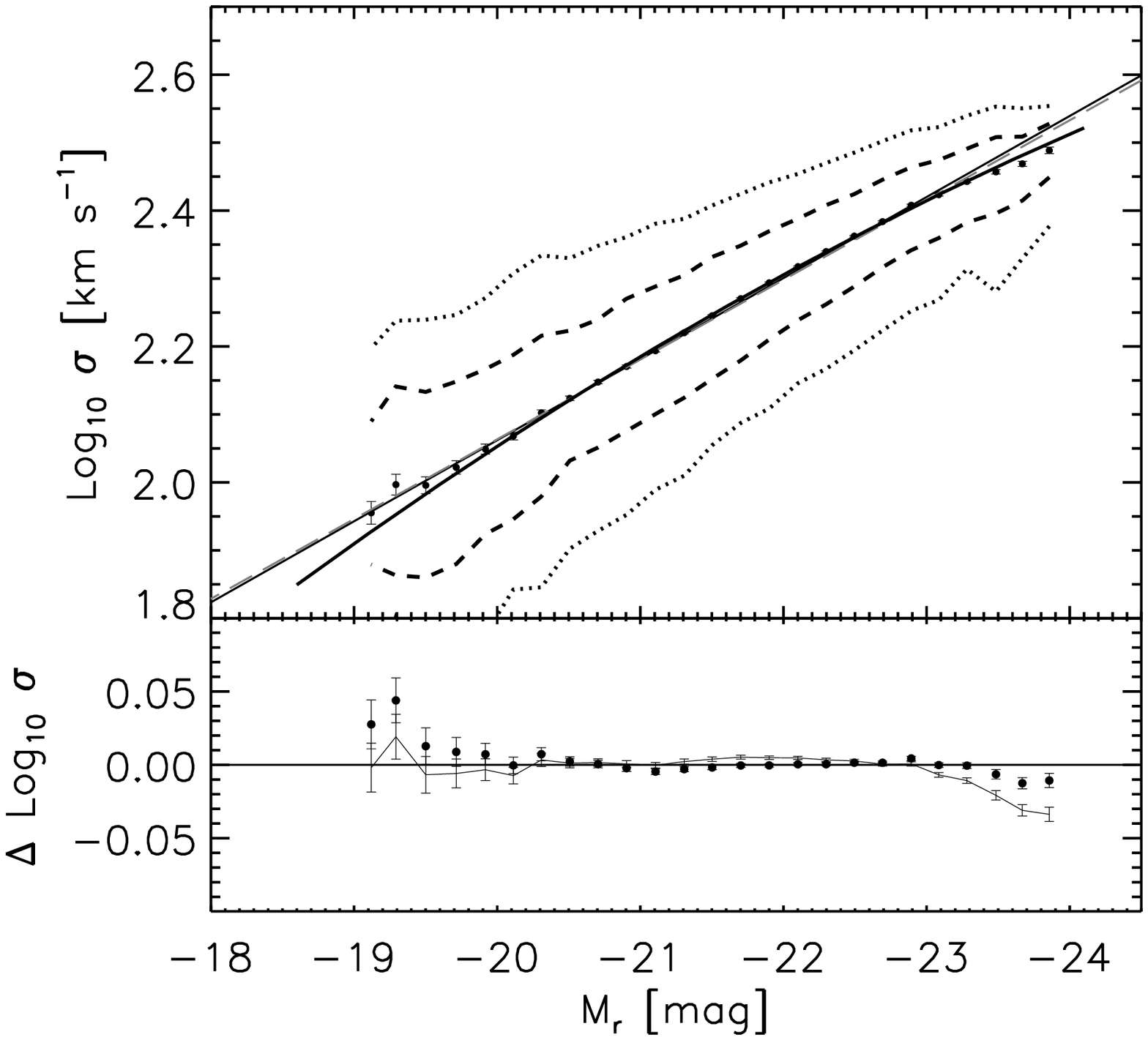}
 \caption{Same as previous figure, but for velocity dispersion.}
 \label{SL}
\end{figure*}

\begin{figure}
 \vspace{-2cm}
 \centering
 \includegraphics[width=\hsize]{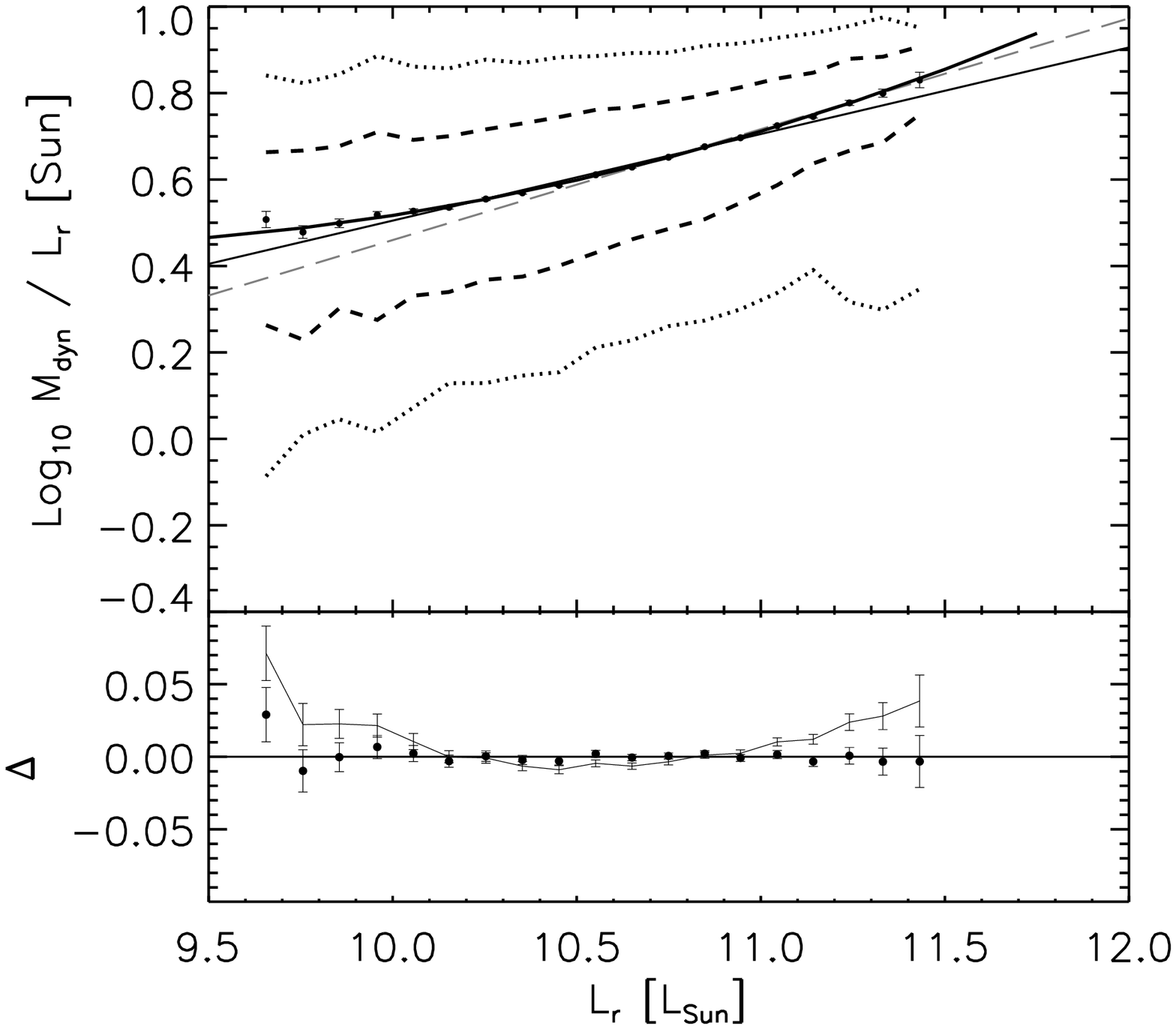}\\
 \vspace{-2cm}
 \includegraphics[width=\hsize]{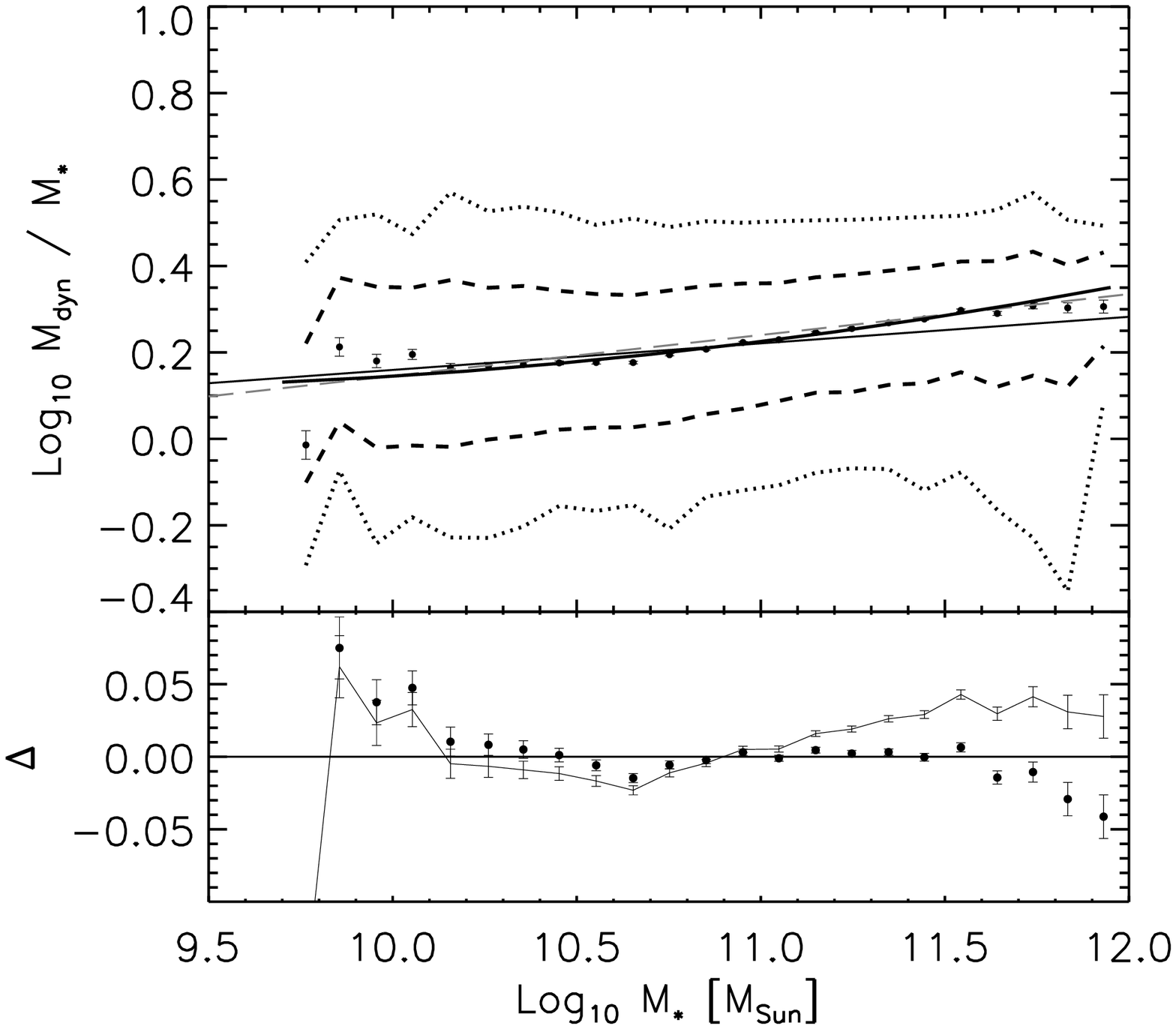}
 \includegraphics[width=\hsize]{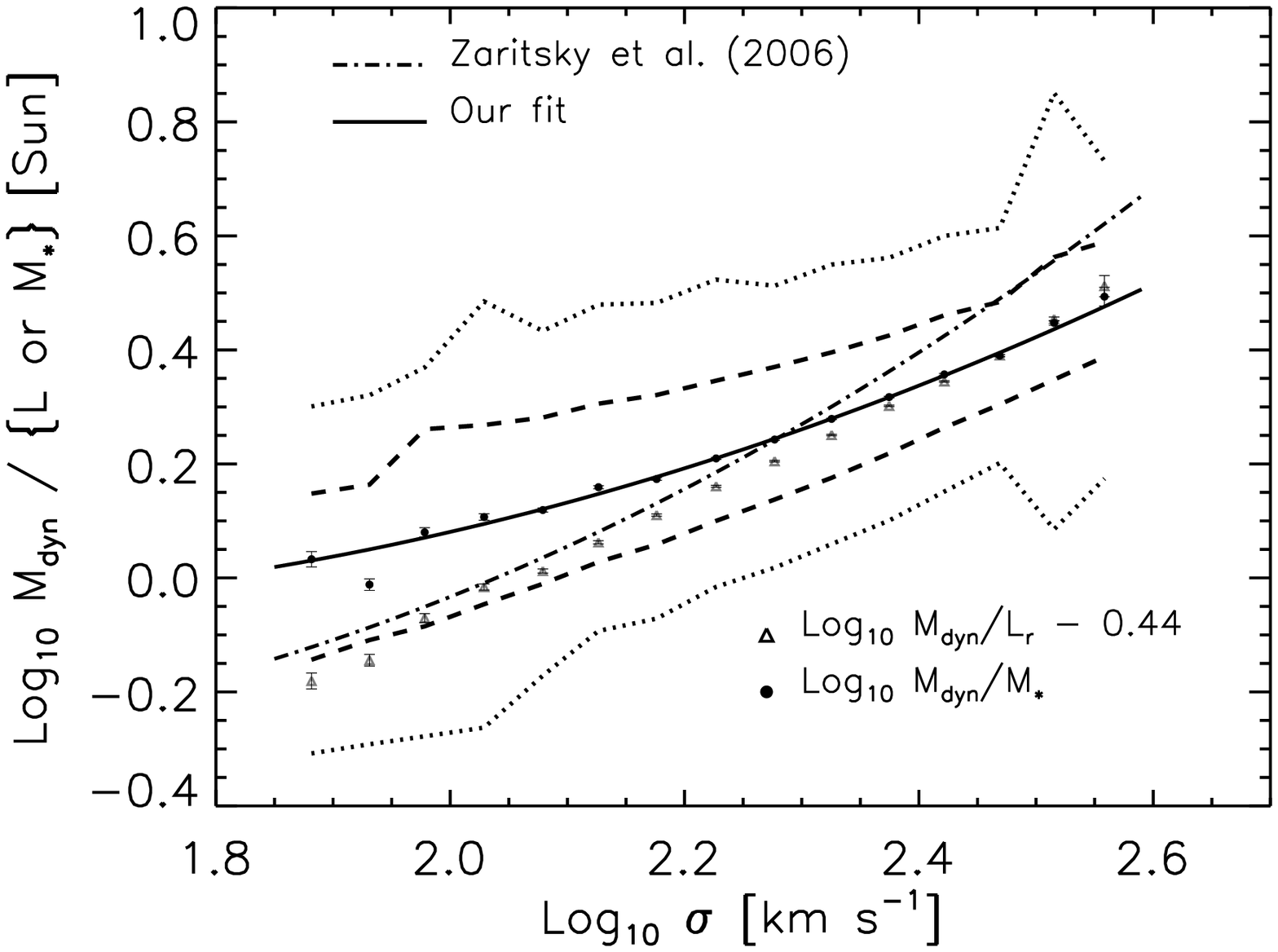}
  \caption{As for previous figure, but now for the ratio of 
          dynamical mass to light versus luminosity (top) 
          and dynamical to stellar mass versus $M_*$ (middle) 
          and $M_{\rm dyn}/L$ or $M_{\rm dyn}/M_*$ versus 
          $\sigma$ (bottom). }
 \label{MdMs}
\end{figure}

A simple test for curvature is as follows.  
In a magnitude limited survey, the more luminous objects are seen 
out to bigger volumes than the fainter objects, meaning that the 
ratio of the number of luminous to faint objects seen in such a 
survey is larger than the ratio of the true number densities of 
these objects.  One can account for this over-abundance by 
weighting each object by $1/V_{\rm max}(L)$, the inverse of 
the volume over which it could have been seen; this down-weights 
luminous galaxies relative to fainter ones.  
If a scaling relation is a straight line, e.g., if the mean 
log(size) increases linearly as log(luminosity) increases, then 
the slope of the regression of $\log R$ on $\log L$ will not 
depend on whether or not one includes this weighting term.  
(This will also be true for $\log\sigma$ on $M$, etc.)  
However, if the intrinsic scaling relation is curved, then the 
slope of the regression line will depend on which weighting was 
used.  This happens to be true in our dataset, indicating that the 
relations are curved:  
the slope of the $\log R -M$ relation is $-0.221\pm0.001$ when 
weighting by $V_{\rm max}^{-1}$, but $-0.241$ when not.  These 
numbers are $-0.119\pm0.001$ and $-0.117$ for the 
$\log\sigma-M$ relation.  
The difference is more dramatic for the $\mu_e-M$ relation:  
$-0.104\pm0.004$ and $-0.207$.  
Here, we reported the uncertainties due to random errors; 
uncertainties in the parameters due to systematics errors 
are a few times larger (see Section~\ref{RL}).

\subsection{The size-luminosity relation}\label{RL}
Recent work shows that the correlation between size and luminosity 
relation has evolved significantly since $z=2$ (e.g. Cimatti et al. 2008; 
van Dokkum et al. 2008).
Bernardi (2009) shows that the sizes of luminous early-type 
galaxies ($M_r \sim -23$~mag) are still evolving at low redshift and 
that satellites are on average $\sim 8 \%$ smaller than central galaxies.
However, this last result is not seen for lower luminosity early-type galaxies
(e.g. Weinmann et al. 2009).    
Due to this recent interest, we study this relation first, before considering 
others.

The symbols in the panels on the right of 
Figures \ref{RL_f} and \ref{SL} show the curvature in the size-luminosity and 
$\sigma$-luminosity relations, respectively (e.g. Lauer et al. 2007; 
Bernardi et al. 2007).  
This curvature is also evident when shown as a function of stellar 
mass (panels on the left), so curvature in the $M_*/L$ relation is 
not the only cause.  

The format of this figure is similar to many that follow.  
The filled circles in the top panel show the median size in a number 
of narrow bins in luminosity, when the objects are weighted by 
$V_{\rm max}^{-1}$.  
Dashed and dotted curves show the regions which include 68\% 
and 95\% of the weighted counts.  
For statistics at fixed luminosity (as in this case), the 
$V_{\rm max}$ weighting makes no difference because, at fixed 
luminosity, all galaxies have the same weight.  
In Figure~\ref{RVIR}, which shows the $R-\sigma$ and $R-\mu_e$ 
relations, the difference is significant.  
There, we use open squares to show the median weighted count when 
no $V_{\rm max}$ weighting is used.  

Straight solid and dashed lines show the result of fitting straight 
lines to the data, with and without $V_{\rm max}$ weighting.  For 
distributions at fixed luminosity (as in this case), the two should 
be the same if the underlying scaling relation is not curved.  If 
it is, then the fit to the unweighted points will reflect the slope 
of the relation at higher luminosities.  In the case of the 
size-luminosity relation (Figure 8), the 
dashed line is steeper than the solid, consistent with the steepening 
of the relation at large $L$ shown by the symbols.  
In this case only, we also show the linear fits to the $R-L$ and 
$R-M_*$ relations reported by Shen et al. (2003)
(see their Table~1, Fig.~6 and Table~1, Fig.~11 respectively). 
This shows that their fits are close to ours when we ignore the 
$V_{\rm max}$ weighting, but note that their sample is selected 
rather differently, and their sizes are from Sersic, rather than 
deVaucouleur fits to the light profile.  

To quantify the curvature, we fit 2nd order polynomials to these 
and a number of other scaling relations (which follow).  In all 
cases our fits are slightly non-standard because, to emphasize the 
curvature in these relations, we would like the fits to be sensitive 
to the tails of the distribution.  
Therefore, we have fit to the binned counts (i.e., the symbols) 
shown in the Figures, rather than to the objects themselves.  
For fits at fixed $L$, this is equivalent to weighting each object, 
not just by $V_{\rm max}^{-1}(L)$, but by
 $[V_{\rm max}(L)\phi(L)]^{-1} = [N_{\rm obs}(L)]^{-1}$.  
In effect, this upweights the tails of the distribution.  
The fitting minimizes $\chi^2$, defined as the sum of the 
squared distances to the binned points shown.  If, instead, 
we weighted each of the binned points by (the inverse of its) 
error bar when defining $\chi^2$, then, because this additional 
weight is proportional to the number of counts in the bin, 
this would be the equivalent of weighting each object, 
rather than each binned point, equally.

The bottom panel shows residuals from the linear 
($V_{\rm max}^{-1}$ weighted) and quadratic fits; in all cases, 
the residuals from the quadratic fits show fewer, if any, trends. 
Table~1 summarizes the results of these fits.  
Although the coefficients of the quadratic term appear to be very 
different from those when simply fitting a line, this is because 
we are reporting fits of the form 
 $\langle Y|X\rangle = p_0 + p_1\, X + p_2\, X^2$.
Had we removed the mean values first, and fit 
 $\langle y|x\rangle = p_0 + p_1\, x + p_2\, [x^2 - \langle x^2\rangle]$
instead (where $x = X-\langle X\rangle$, etc.)
then $p_1$ would be very similar to linear fit, indicating that 
the curvature is small.  

To quantify if a quadratic is a significantly better fit than a 
straight line, we fit both to the binned counts.  We then compare 
 $\chi^2_{\rm quad}/(N_{\rm bins}-3)$ with 
 $\chi^2_{\rm line}/(N_{\rm bins}-2)$, 
where $\chi^2_{\rm quad}$ and $\chi^2_{\rm line}$ denote the 
minimized values of $\chi^2$, and the fits were to $N_{\rm bins}$ 
binned points.  
In all cases, this reduced value of $\chi^2$ for the quadratic 
is much closer to unity than for the linear fits.  
(Note that we do not use the coefficients of the linear fits to 
the unbinned counts that are shown in Table~1, since, for this 
test, we want to treat both the linear and quadratic fits equally.  
Of course, using these coefficients when computing 
$\chi^2_{\rm line}$ does not change our conclusions, since they 
can, at best, produce the same value of $\chi^2_{\rm line}$ 
which we have just described.)   

Finally, a word on the errors on the fitted coefficients is in 
order.  The numbers we quote are random errors:  they depend on 
slope of the relation, its scatter, and the sample size.  These 
are smaller than systematic effects:  e.g., using $\sigma_{\rm DR6}$ 
or $\sigma_{\rm IDLspec2d}$ rather than their average makes a 
difference which is larger than the random error.  
Typical systematics errors are a few times larger than random errors --  
therefore, it is important to separate the two types of error.

Figure~\ref{RL_f} shows that, compared to a single power law, 
the $R-L$ relation curves significantly towards larger sizes at 
large $L$ or $M_*$, consistent with previous work.  
At small $L$ or $M_*$ ($M_r<-19.5$ or $\log_{10} M_*/M_\odot < 10.2$), 
the data appear to scatter slightly downwards from the quadratic fit.  
Although there are few objects in this tail, so the measurement is 
noisier, it is possible that this indicates that the sample is 
slightly contaminated at small $L$.  We will return to this shortly.  

Before we move on to other scaling relations, we note that results 
based on Petrosian quantities are shown in the Appendix, where we 
also discuss why we do not consider them further.  

\begin{figure*}
 \vspace{-2cm}
 \centering
 \includegraphics[width=0.475\hsize]{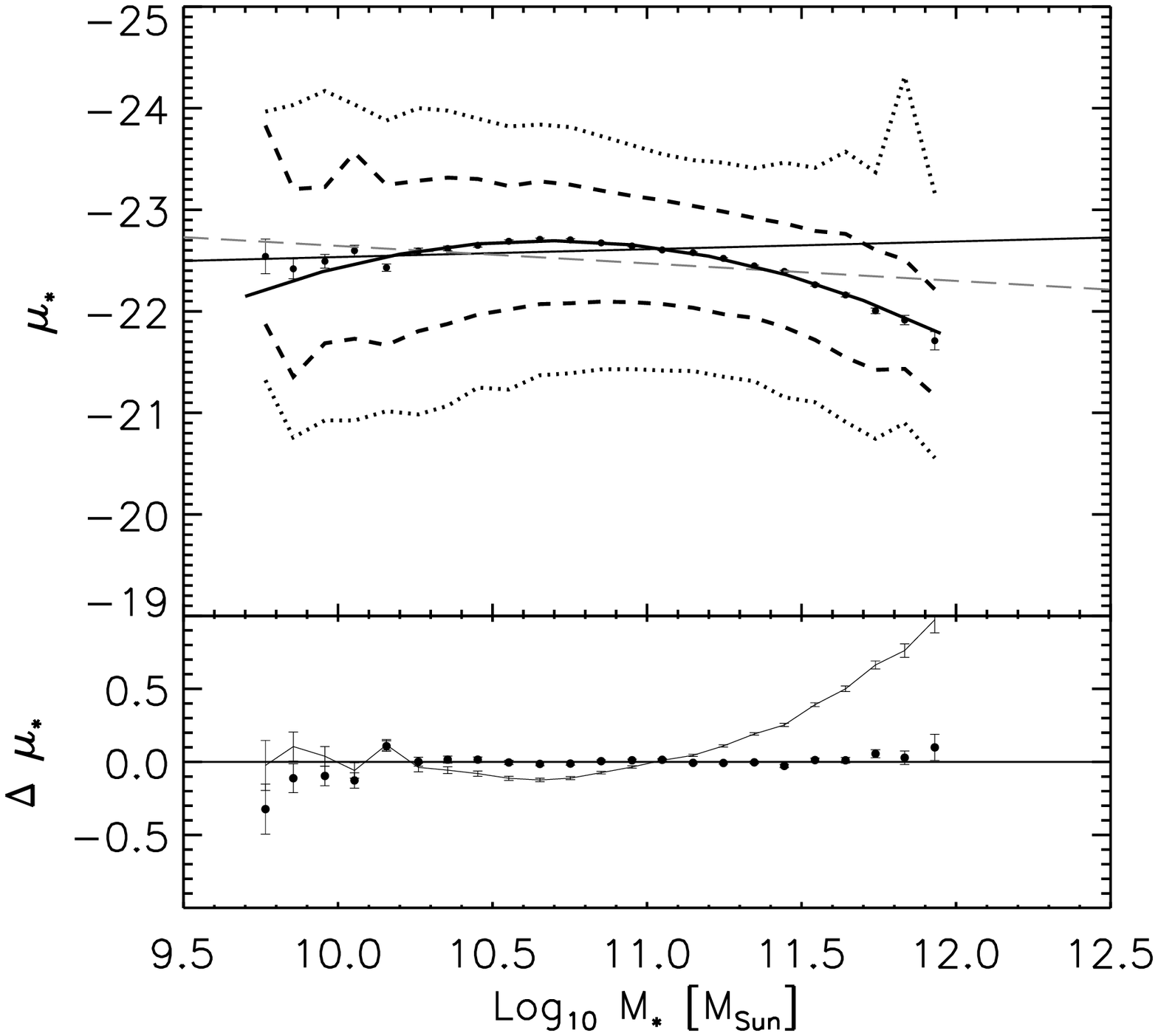}
 \includegraphics[width=0.475\hsize]{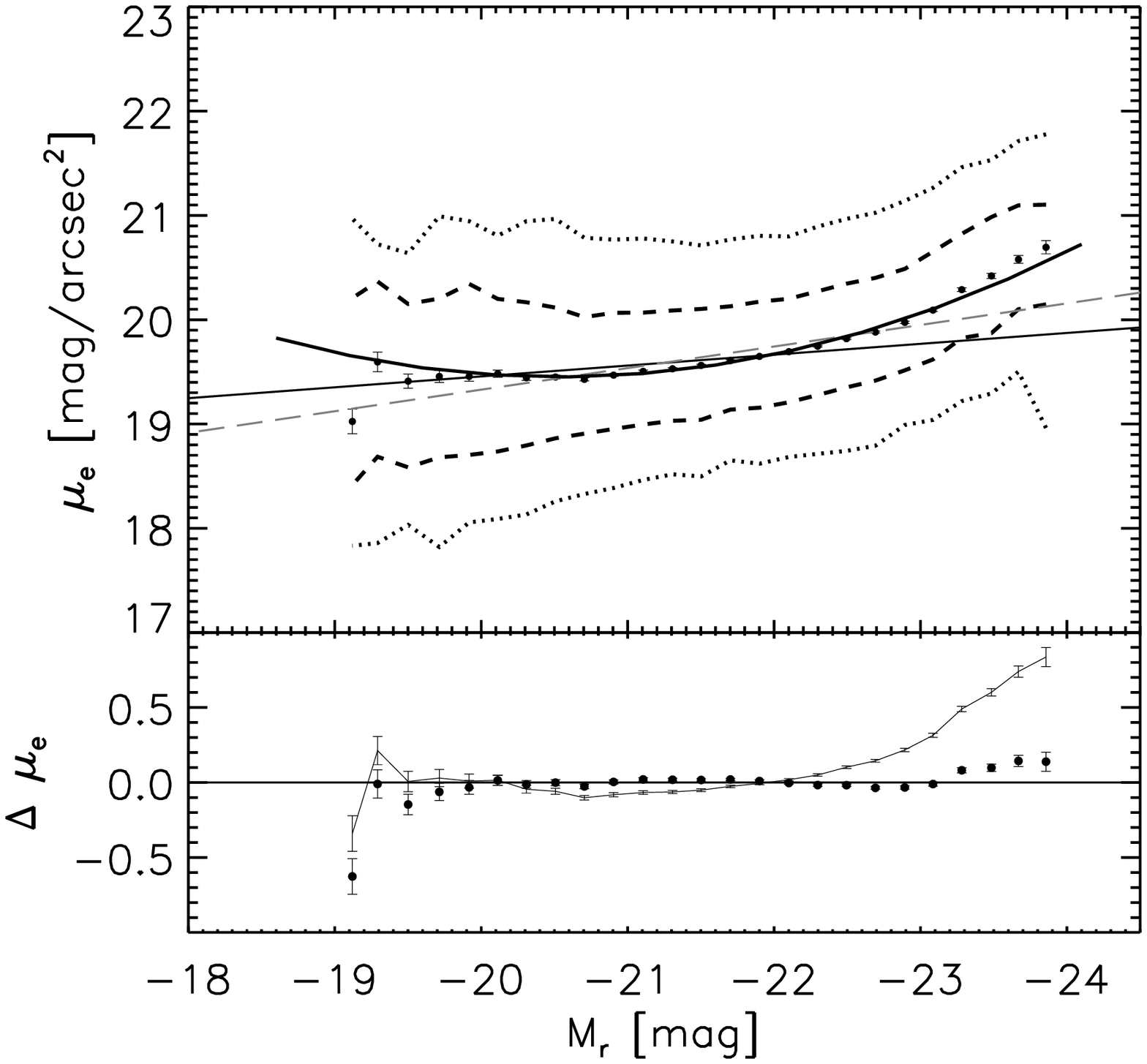}
 \caption{As for previous figure, but for stellar mass surface-brightness vs 
          stellar mass (left) and surface-brightness vs luminosity (right).  }
 \label{Mus}
\end{figure*}

\subsection{Other scaling relations}
Figure~\ref{SL} shows the $\sigma-L$ relation in this sample.  
This relation is actually rather well described by a single 
power-law, except at $M_r<-23$ (where the mean value 
of $\log_{10}(\sigma/{\rm km~s^{-1}}) > 2.4$) where it curves 
slightly downwards.  The flattening of the $\sigma-L$ relation 
is consistent with previous work, but notice that this effect 
is even more pronounced for the $\sigma-M_*$ relation.

\begin{figure*}
 \vspace{-2cm}
 \centering
 \includegraphics[width=0.475\hsize]{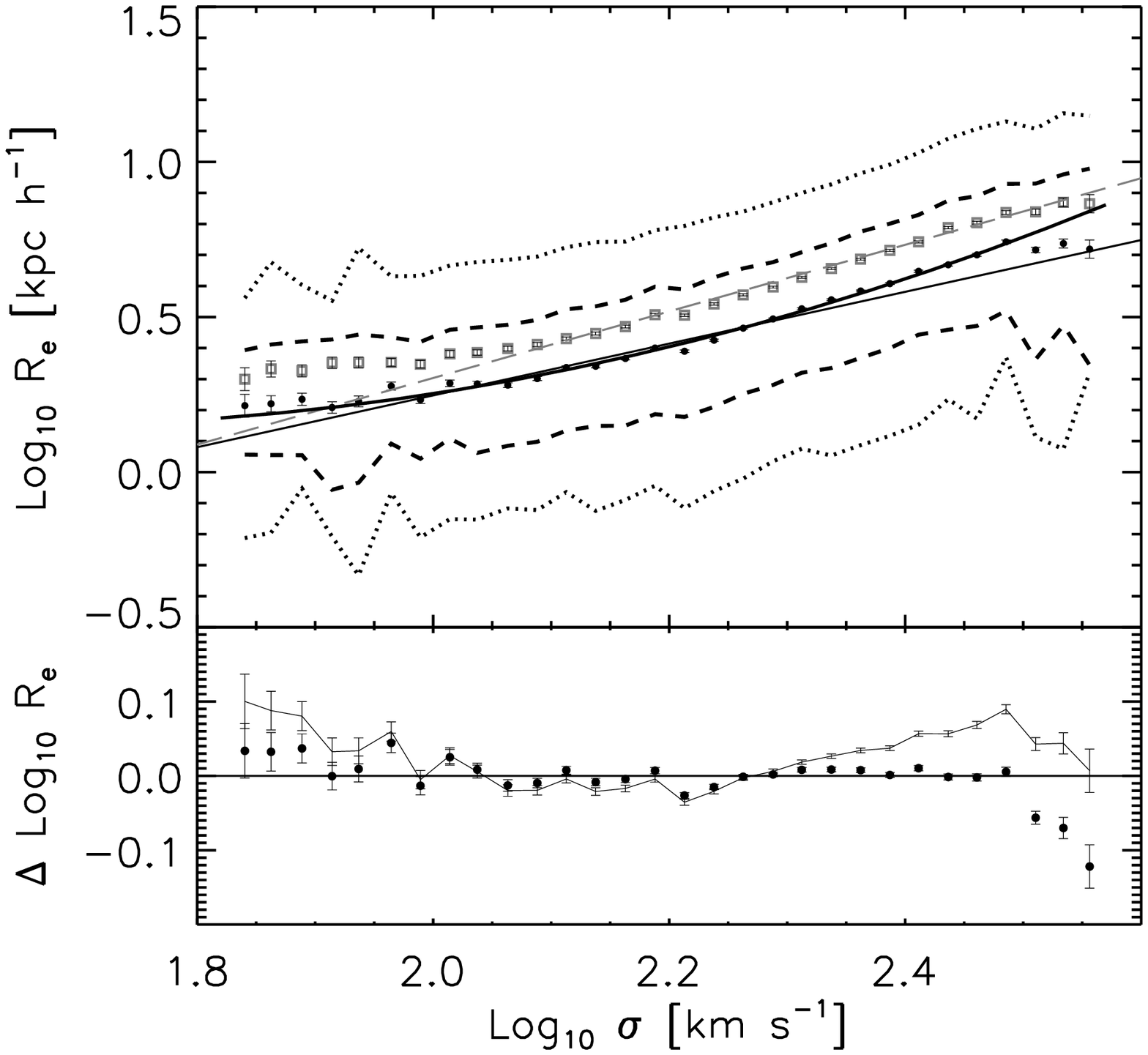}
 \includegraphics[width=0.475\hsize]{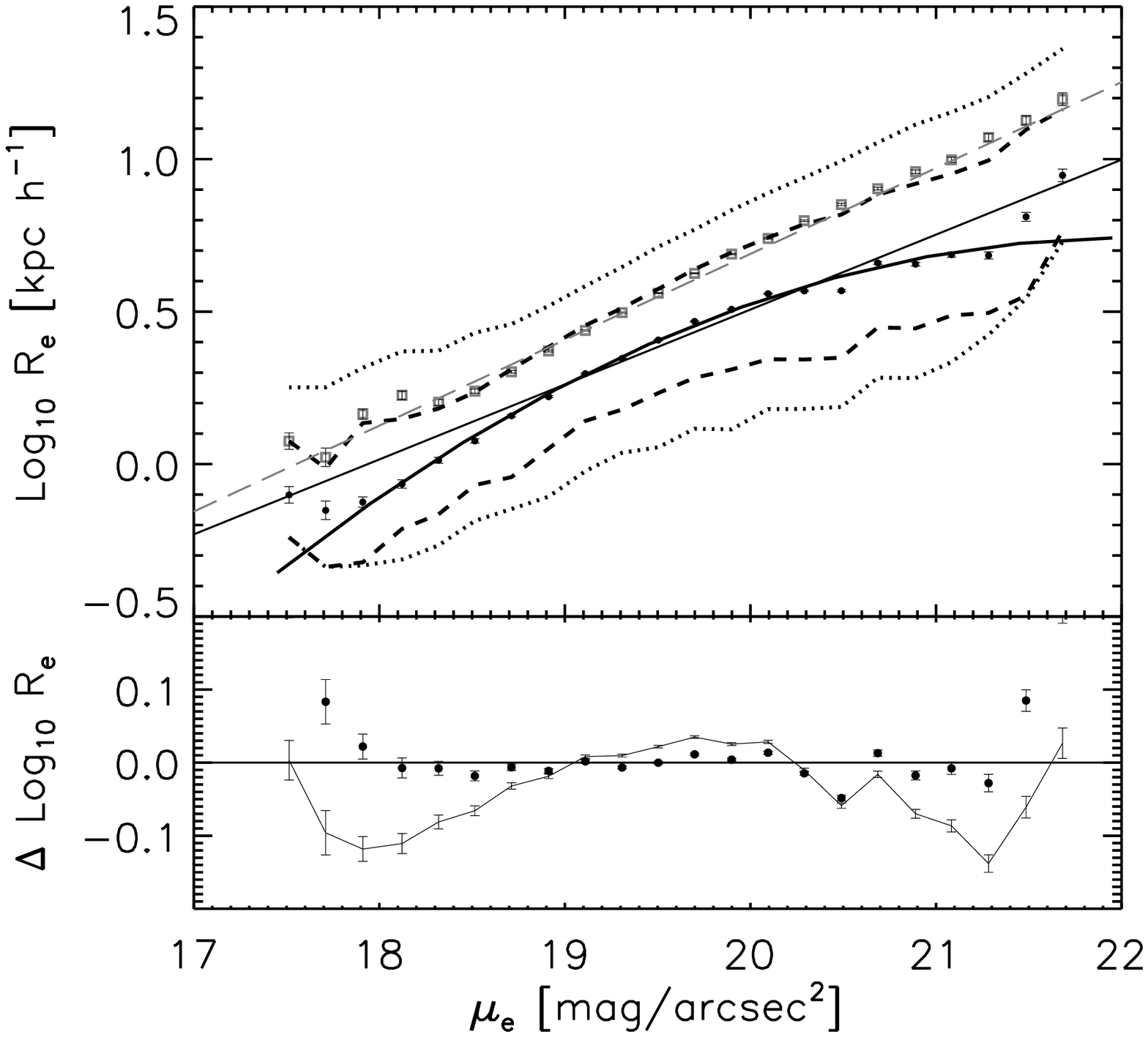}
 \caption{As for previous figure, but now for the size-velocity 
          dispersion and size-surface-brightness relations.}
 \label{RVIR}
\end{figure*}

Comparison with Figure~\ref{RL_f} shows that, at large $M_*$ and $L$, 
the $R-L$ and $\sigma-L$ relations curve in opposite senses. 
Since dynamical mass $M_{\rm dyn}\propto R\sigma^2$, it is 
interesting to check if the $M_{\rm dyn}-L$ relation is well-fit 
by a simple power-law.  We do this in Figure~\ref{MdMs}, where 
we set 
 $M_{\rm dyn} \equiv 5\,R_e\sigma^2/G = 4.65\times 10^{10}h^{-1}M_\odot 
  \,(R_e/h^{-1}{\rm kpc})\,(\sigma/200~{\rm km~s}^{-1})^2$.  
The top panel shows that, at both small and large $L$, the 
$M_{\rm dyn}-L$ relation curves upwards from the 
$R\sigma^2\propto L^{0.2}$ scaling (solid line).
The middle panel shows that curvature remains if one replaces 
$L$ with $M_*$, indicating that more than stellar population 
related effects are responsible.  
The slight rise at small masses is not implausible: star formation 
is expected to be less efficient at small masses.  However, we view 
this with caution:  the velocity dispersions at the small-mass end 
are more uncertain (e.g. Bernardi 2007). 
There is an average trend for $M_{\rm dyn}/M_*$ 
to increase with $M_*$, although it is weak: 
$\langle M_{dyn}/M_*|M_{dyn}\rangle \propto M_{*}^{0.062\pm 0.006}$
 (the error 0.06 on the slope was computed accounting for systematics 
errors -- the uncertainty from random errors is smaller $\sim 0.002$).
The correlation between $M_{\rm dyn}/M_*$ with $M_{\rm dyn}$ is stronger:
$\langle M_{dyn}/M_*|M_{dyn}\rangle \propto M_{dyn}^{0.17\pm 0.01}$
(see Figure~11 in Hyde \& Bernardi 2008).
If there were no scatter around this relation, then we would expect 
 $\langle M_{dyn}/M_*|M_*\rangle \propto M_*^{0.17/0.83}\propto M_*^{0.2}$; 
because there is scatter, this scaling is shallower, 
$\propto M_*^{0.06\pm0.01}$.
Even if this relation were 
a simple power law without curvature, it would indicate that stars make 
up a smaller fraction of the total mass of a galaxy at large masses.  This 
provides an important piece of information for adiabatic contraction 
based models of scaling laws (e.g. Padmanabhan et al. 2004; 
Lintott et al. 2005).

To connect with previous work, the bottom panel shows how 
$M_{\rm dyn}/L$ (triangles) and $M_{\rm dyn}/M_*$ (filled circles) 
scale with $\sigma$; both relations are slightly curved.  Except at 
the largest $\sigma$, our data are relatively well described by the curved 
$M_{\rm dyn}/L - \sigma$ relation reported by Zaritsky et al. (2006) 
(we have shifted our measurements downwards by 0.44~dex because 
our data are in $r$ whereas their fit was in $I$, i.e. we use $r-I=1.1$.
Note that we have also subtracted $-0.3$ from their fit since they used 
the effective light $L_e = L/2$, while we use the total light $L$).  
The fact that $M_{\rm dyn}/L$ is a steeper function of $\sigma$ 
than is $M_{\rm dyn}/M_*$ can be understood as follows.  
First, note that $M_{\rm dyn}/L = (M_{\rm dyn}/M_*)\,(M_*/L)$.  
Then, note that $M_*/L$ increases with increasing $g-r$ color 
(e.g. Bell et al. 2004).  However, $g-r$ color is strongly 
correlated with $\sigma$ -- large $\sigma$ implies redder colors  
(Bernardi et al. 2005), and so $M_*/L$ increases with $\sigma$.  
Therefore, $M_{\rm dyn}/L$ increases with $\sigma$ because 
$M_{\rm dyn}/M_*$ does and because $M_*/L$ does so as well.

As a final study of curvature in relations which involve luminosity, 
we now turn to the $\mu_e-L$ and $\mu_*-M_*$ relations.  (We define 
 $\mu_* = -2.5\log_{10}(M_*/M_\odot)+ 5\log_{10}(R_e/{\rm kpc}) 
          + 2.5\log_{10}(2\pi)$;
this is the stellar mass surface brightness within the half 
light radius.)  
Figure~\ref{Mus} shows that in this case too, there is significant 
curvature.  
However, the panel on the left shows that at 
 $M_* < 3\times 10^{10}M_\odot$, 
the $\mu_*-M_*$ relation becomes rather well fit by the linear 
relation, although it also becomes significantly noisier.  
A look back at the other relations which involve $M_*$ shows that 
they too become less well-defined at small $M_*$.
This happens to be the same mass scale which 	Kauffmann et al. (2003) 
identify as being special.  

It is possible that our early-type sample is contaminated by a 
different population at the low mass end -- despite the fact that 
we have already tried to reduce such an effect by removing objects 
with $b/a < 0.6$ and selecting galaxies with the $g$ and $r$-band
{\tt fracDev} $= 1$. If we include those objects with $g$ and $r$-band 
{\tt fracDev} $> 0.8$ and/or do not remove objects with small $b/a$,
then the quadratic remains a good fit even at small $M_*$ or $L$.  
This is also true if we simply use the cuts given by 
Kauffmann et al. (2003):  ${\tt R90/R50} > 2.86$, ${\tt R50}>1.6$ 
and $\mu_{50} < 23$~mag$/$arcsec$^2$ in the $r$-band.  
We will return to this in the next subsection, but note that the 
curvature at the luminous, massive end of the sample, is highly 
significant.

We turn now to the scaling relations which play a fundamental 
role in the Fundamental Plane: the $R_e-\sigma$ and $R_e-\mu_e$ 
relations.  Figure~\ref{RVIR} shows that both these relations 
are curved.  Notice that accounting for selection 
effects is important -- the filled and empty symbols, 
which include or ignore the $V^{-1}_{\rm max}$ weight, 
trace very different relations.  
To first order, the zero-points of the 
two relations are more strongly affected than is the slope.  
Since it is the zero-point of the FP which is used to estimate 
evolution in small high-redshift samples, Figure~\ref{RVIR} 
suggests that, without due care, one may simply be measuring 
selection effects (a point also made by Bernardi et al. 2003c).  

Neither of these relations is as well-fit by a quadratic at the extremes; 
e.g., when weighted by $V^{-1}_{\rm max}$, the 
$R_e-\mu_e$ relation curves away significantly from the quadratic 
at both large and small $\mu_e$ (see also Nigoche-Netro et al. 2008).  
Curvature at large $\mu_e$ (or $\sigma$) 
is not surprising: we already know that BCGs, and more generally, 
large mass galaxies (Bernardi et al. 2008), follow different scaling 
relations.  
The flattening at small $\sigma$ is perhaps more surprising; 
this is in the regime where the SDSS velocity dispersions are 
most suspect, so one might worry that some of the effect is from 
measurement errors scattering objects to smaller $\sigma$.  However, 
there is a similar flattening of the $R_e-\mu_e$ relation at $\mu_e<18$.  
What causes this?

\subsection{Curvature from contamination?}\label{contam}

\begin{figure}
 \centering
 \includegraphics[width=\hsize]{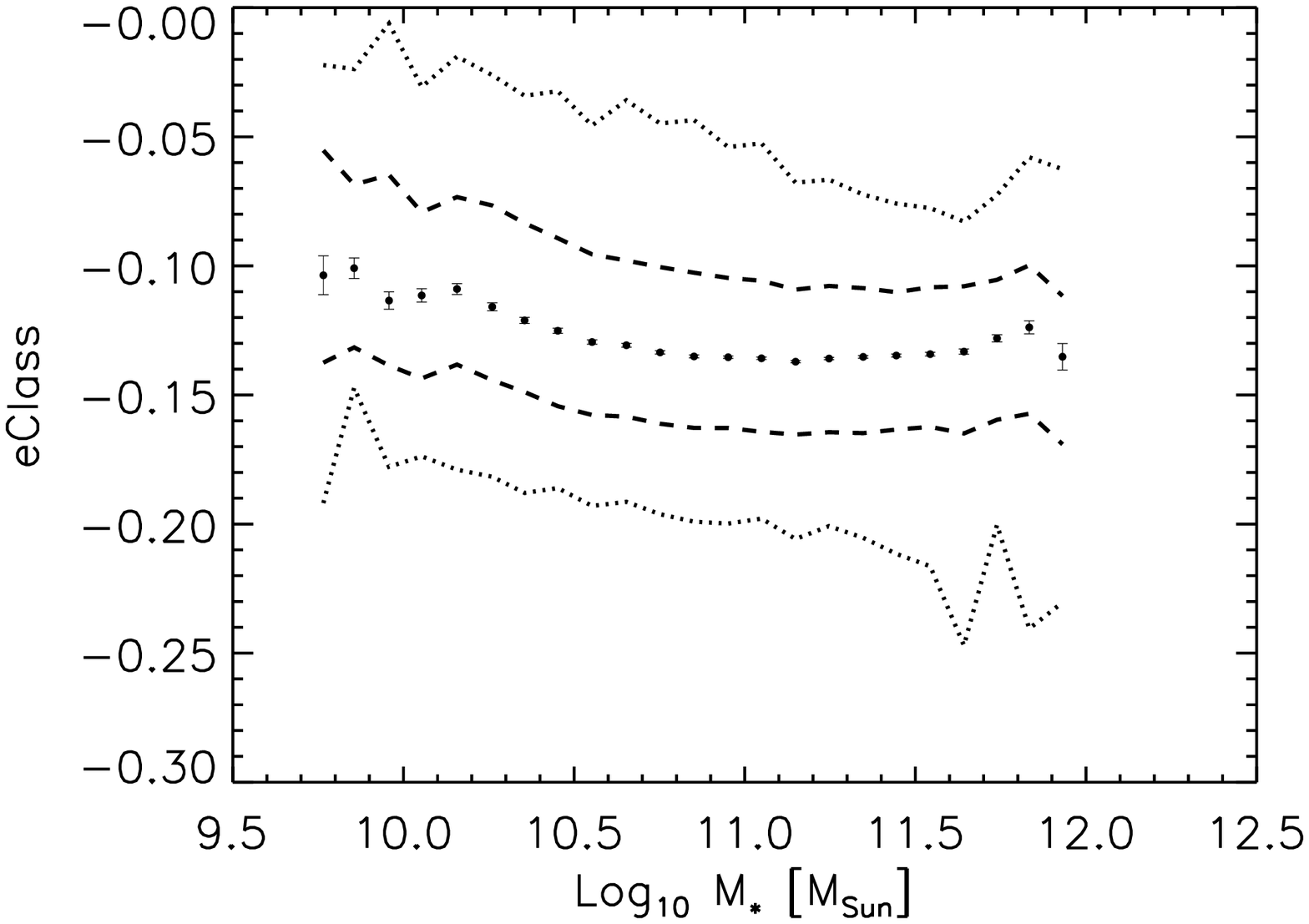}
 \includegraphics[width=\hsize]{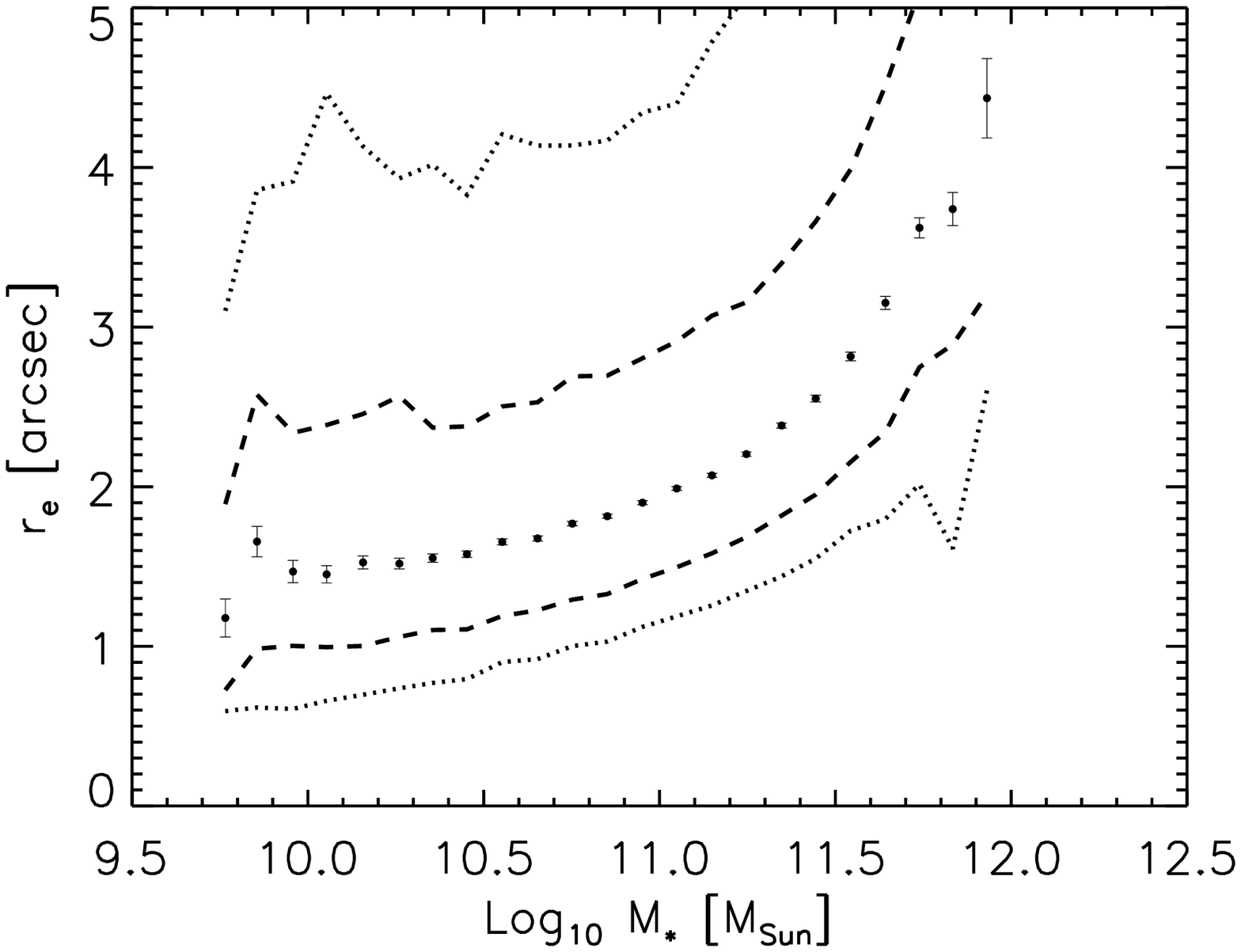}
 \caption{PCA eClass (top) and angular size (bottom) as a function 
          of stellar mass.  Later-type galaxies have less negative
          eClass values.}
 \label{eclassM*}
\end{figure}

\begin{figure}
 \vspace{-2cm}
 \centering
 \includegraphics[width=0.95\hsize]{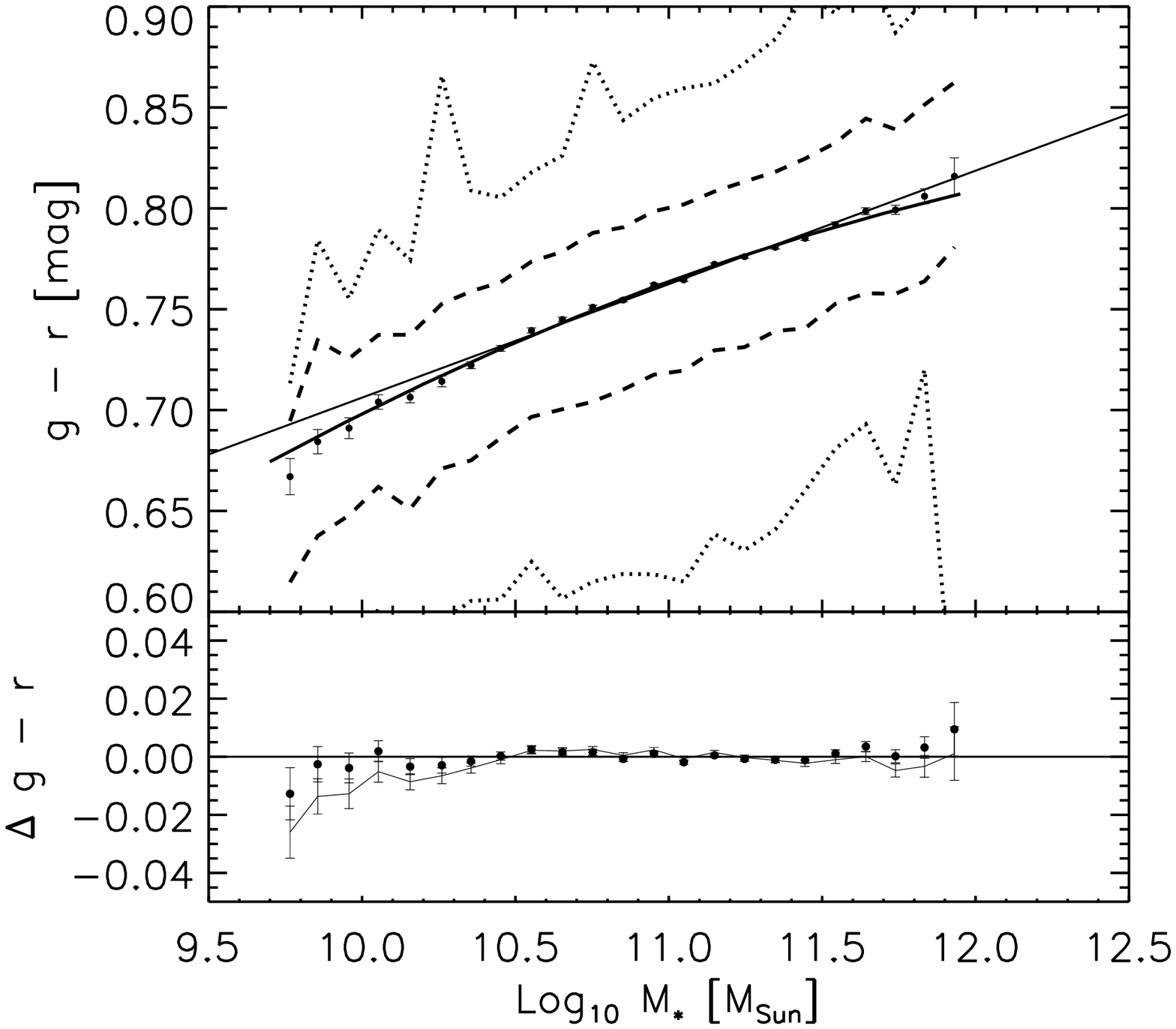}\\
 \vspace{-1.8cm}
 \includegraphics[width=0.95\hsize]{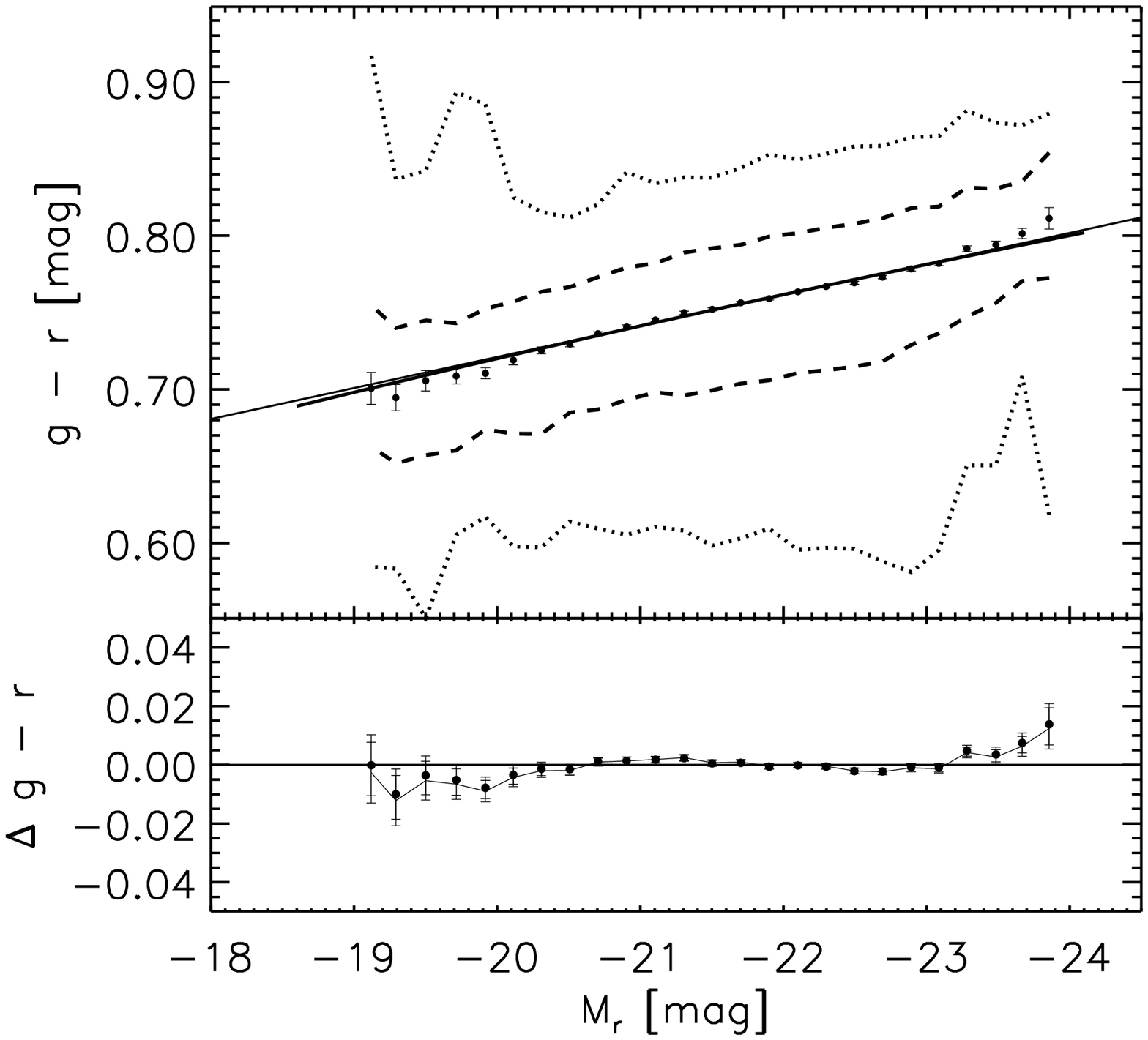}
 \caption{Similar to previous figures, but for the color-magnitude 
          and color-stellar mass relations.}
 \label{CM}
\end{figure}

In the previous subsection we raised the possibility that the
sample may be contaminated at low $M_*$.  With this in mind, we 
performed a visual inspection of the objects with small $M_*$; 
this did not reveal any peculiarities.  
However, the top panel of Figure~\ref{eclassM*} shows the results 
of a PCA analysis of the spectra.  A small departure from early- 
towards later-type values is seen at small-$M_*$, but note that 
true late-types have eClass values which are greater than zero.  
If we assume that some fraction $f$ of the sample has eClass = 0, 
and the rest has eClass = $-0.13$, then to find eClass = $-0.11$ 
requires $f\approx 0.15$.  The true contamination is likely to 
be smaller (since eClass $>0$ for late-types).  

Alternatively, one might have worried that this is an aperture 
effect associated with abundance gradients -- the spectra are 
from fibers which take light from a fixed angular radius, so, for 
smaller galaxies, the light from the inner bulge contributes a 
decreasing fraction of the total light in the fiber.  However, a 
plot of $r_e$ vs $M_*$ is approximately constant at small $M_*$ 
(bottom panel of Figure~\ref{eclassM*}), suggesting that aperture 
effects are not to blame.  

Finally, we note that the color-stellar mass and color-magnitude
relations also show a small curvature, towards bluer colors, at small $L$ 
or $M_*$.  The curvature in these relations is enhanced if we 
allow the full range of $b/a$ in our sample (rather than removing 
objects with $b/a<0.6$) and if objects with $g$ and $r$-band 
{\tt fracDev} $> 0.8$ are included (rather than selecting galaxies 
with {\tt fracDev} $= 1$). This, further circumstantial evidence 
for a small amount of contamination, is shown in Figure~\ref{CM}.  

\begin{figure*}
 \vspace{-2cm}
 \centering
 \includegraphics[width=0.475\hsize]{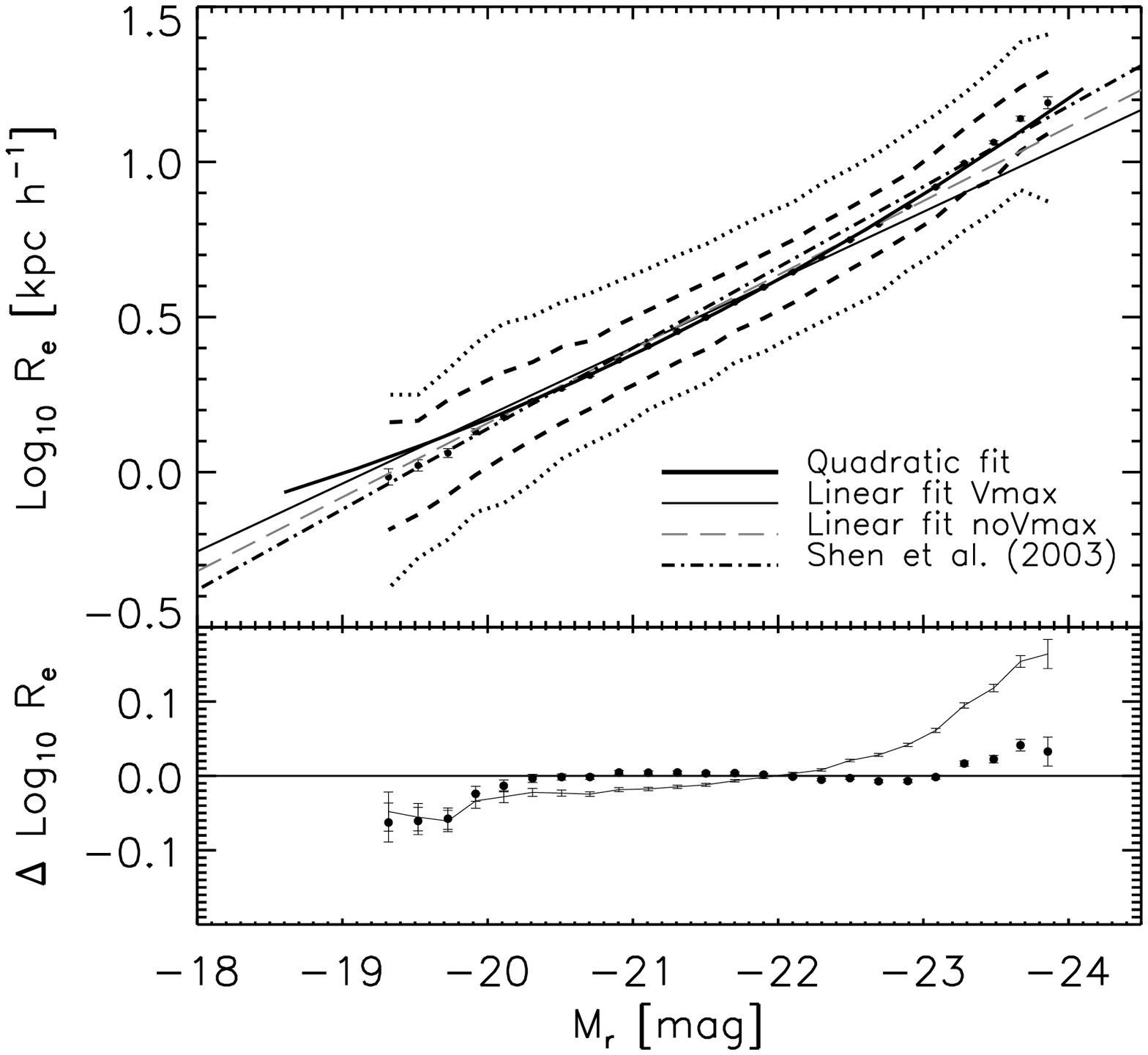}
 \includegraphics[width=0.475\hsize]{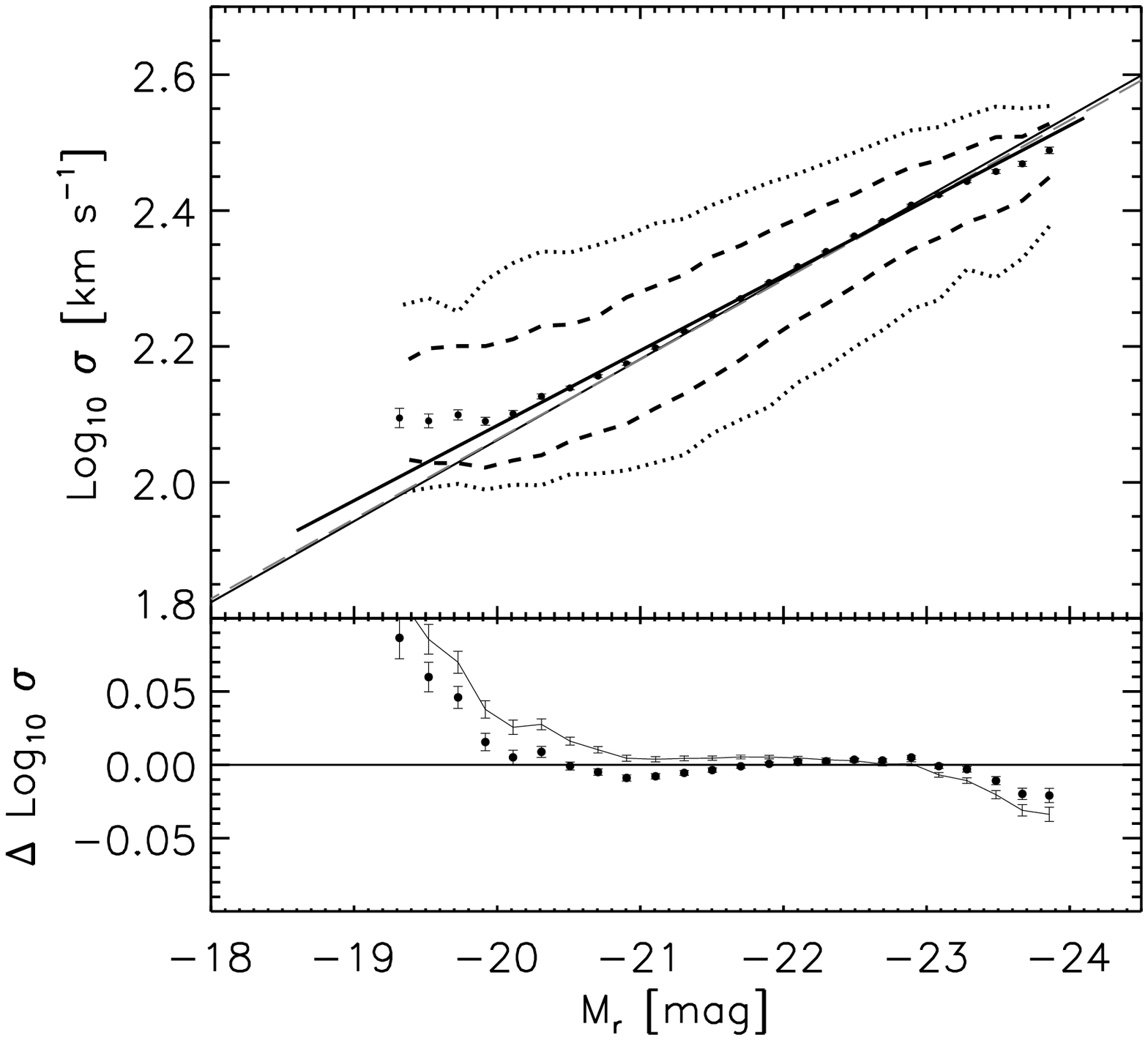}\\
 \vspace{-1.8cm}
 \includegraphics[width=0.475\hsize]{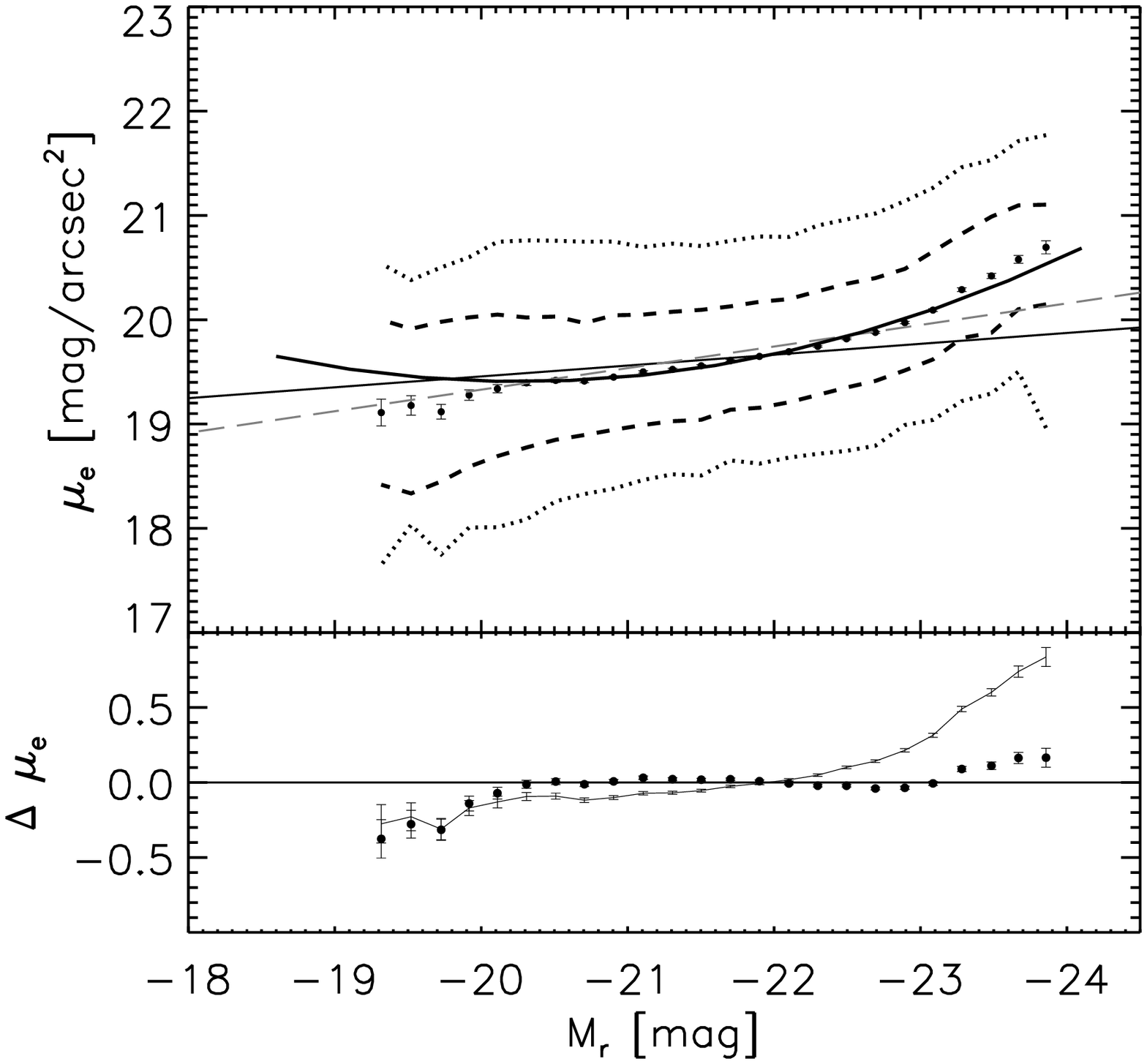}
 \includegraphics[width=0.475\hsize]{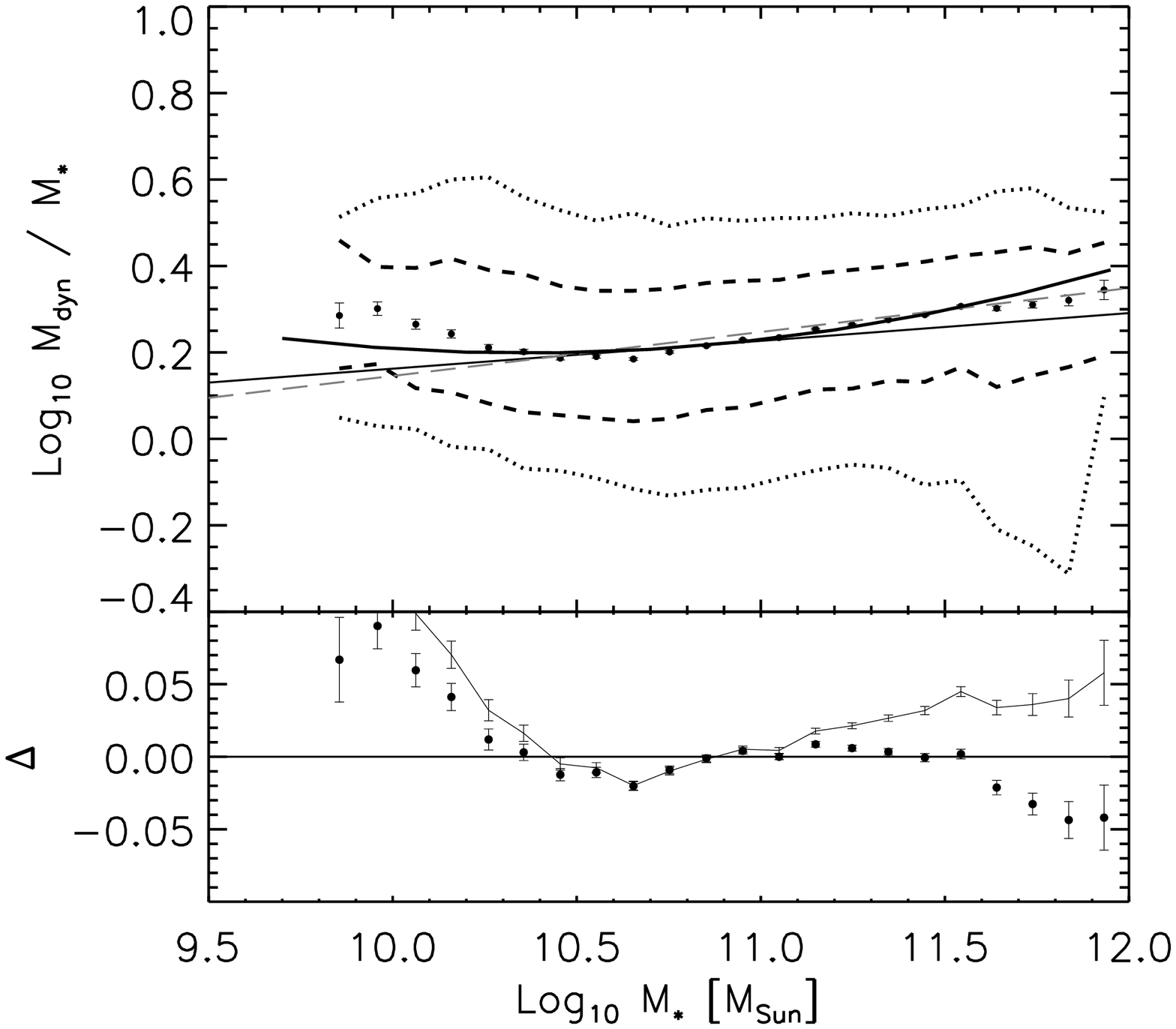}\\
 \caption{Effect on scaling relations of removing objects with 
          $\sigma<90$~km~s$^{-1}$.  
          Smooth solid lines show the linear fits to the 
          full sample (taken from previous figures); 
          symbols show the scaling relations when objects with 
          $\sigma<90$~km~s$^{-1}$ have been removed; dashed and 
          dotted lines enclose 68\% and 95\% of the objects. 
          Parabolas show the quadratic fits to the symbols.}
 \label{sigma90}
\end{figure*}

We have also studied what happens if we remove objects with 
$\sigma < 90$~km~s$^{-1}$.  This is prompted by Figure~\ref{RVIR}, 
and also by the fact that Bernardi et al. (2003a) excluded such 
objects from their study of SDSS early-types (on the grounds that 
this is the regime in which the SDSS dispersions are suspect).  
Figure~\ref{sigma90} shows that removing such objects has a 
dramatic effect on the scaling relations which involve surface 
brightness, since this preferentially removes objects with large 
sizes for their luminosities (or stellar masses). 
As a result, at the small mass end the size-luminosity (stellar mass)
scaling relation leans towards smaller sizes. On the other hand, 
the $\sigma-L$ relation now flattens significantly at small $L$, 
because the small $\sigma$s at small $L$ have been removed (there are 
essentially no objects with small $\sigma$ at large $L$), and the 
luminosity-$\mu_e$ flattens significantly.  
In addition, $M_{\rm dyn}/M_*$ vs $M_*$ becomes more curved because, 
for a given $M_*$, one is removing small $M_{\rm dyn}$.  This leaves 
only large values of $M_{\rm dyn}/M_*$ at small $M_*$, making the 
relation curve upwards steeply.  
See Hyde \& Bernardi (2008) for further discussion of the dramatic 
effects that cuts in $\sigma$ can have on the Fundamental Plane 
(also see Bernardi et al. 2003c).  

\section{Discussion}
We used our own photometric reductions (GALMORPH) of about 6000 
early-type galaxies from the SDSS to calibrate corrections to 
SDSS photometry (equations~\ref{rnew}--\ref{rfit}) which are most 
necessary for photometry of extended objects  
(Figures~\ref{gmorph-sdss} and~\ref{histR}).  
We applied these corrections to a larger sample of about 50000 
early-types, and then analyzed a number of galaxy scaling 
relations in the sample.  Selection of the sample, which is described 
in some detail in Section~\ref{sample}, was more conservative than 
in previous work based on SDSS data.

Small but statistically significant curvature was found for 
all the relations (Figures~\ref{RL_f}--\ref{RVIR}).  
Table~\ref{tab:parabolas} lists the coefficients of best-fitting 
second-order polynomials which provide a concise way of 
describing this curvature.  The Table also provides the 
coefficients of linear fits, for comparison.  
Whereas curvature at large luminosities/stellar masses is 
expected -- BCGs are known to be a special population -- some 
of the scaling relations in our sample show curvature at small 
masses as well.  The critical mass scale is about
 $M_*\approx 3\times 10^{10}\,M_\odot$.  
Whereas we see some evidence from the colors and spectra that 
objects below this mass scale are of slightly later type
(Figures~\ref{eclassM*} and~\ref{CM}), a visual inspection of 
the images shows no obvious peculiarities.  

Our analysis also showed that the ratio of dynamical to stellar 
mass increases at large masses (Figure~\ref{MdMs}); this is a 
useful constraint on models of early-type galaxy formation.  
In addition, the $R-\sigma$ and $R-\mu_e$ relations were shown to 
be rather sensitive to selection effects (Figure~\ref{RVIR}); 
this matters for analysis of the Fundamental Plane.  
The question of whether or not the Fundamental Plane is curved or 
warped is addressed in a companion paper (Hyde \& Bernardi 2008).  
Finally, we showed that seeing biases the scaling relations associated 
with Petrosian-based quantities (Appendix~A), making them ill-suited 
for precision analyses in large ground-based datasets.

\section*{Acknowledgements}
We thank R. K. Sheth for helpful discussions, and M. Sako for generously providing
computing resources. 
J.B.H. was supported in part by a Zaccaeus Daniels fellowship. 
J.B.H. and M.B. are grateful for additional support provided by 
NASA grant LTSA-NNG06GC19G. 

Funding for the Sloan Digital Sky Survey (SDSS) and SDSS-II Archive has been
provided by the Alfred P. Sloan Foundation, the Participating Institutions, the
National Science Foundation, the U.S. Department of Energy, the National
Aeronautics and Space Administration, the Japanese Monbukagakusho, and the Max
Planck Society, and the Higher Education Funding Council for England. The
SDSS Web site is http://www.sdss.org/.

The SDSS is managed by the Astrophysical Research Consortium (ARC) for the
Participating Institutions. The Participating Institutions are the American
Museum of Natural History, Astrophysical Institute Potsdam, University of Basel,
University of Cambridge, Case Western Reserve University, The University of
Chicago, Drexel University, Fermilab, the Institute for Advanced Study, the
Japan Participation Group, The Johns Hopkins University, the Joint Institute
for Nuclear Astrophysics, the Kavli Institute for Particle Astrophysics and
Cosmology, the Korean Scientist Group, the Chinese Academy of Sciences (LAMOST),
Los Alamos National Laboratory, the Max-Planck-Institute for Astronomy (MPIA),
the Max-Planck-Institute for Astrophysics (MPA), New Mexico State University,
Ohio State University, University of Pittsburgh, University of Portsmouth,
Princeton University, the United States Naval Observatory, and the University
of Washington.

\appendix

\section{Petrosian-related scaling relations are biased by seeing}

\begin{figure}
 \centering
 \vspace{-2cm}
 \includegraphics[width=0.95\hsize]{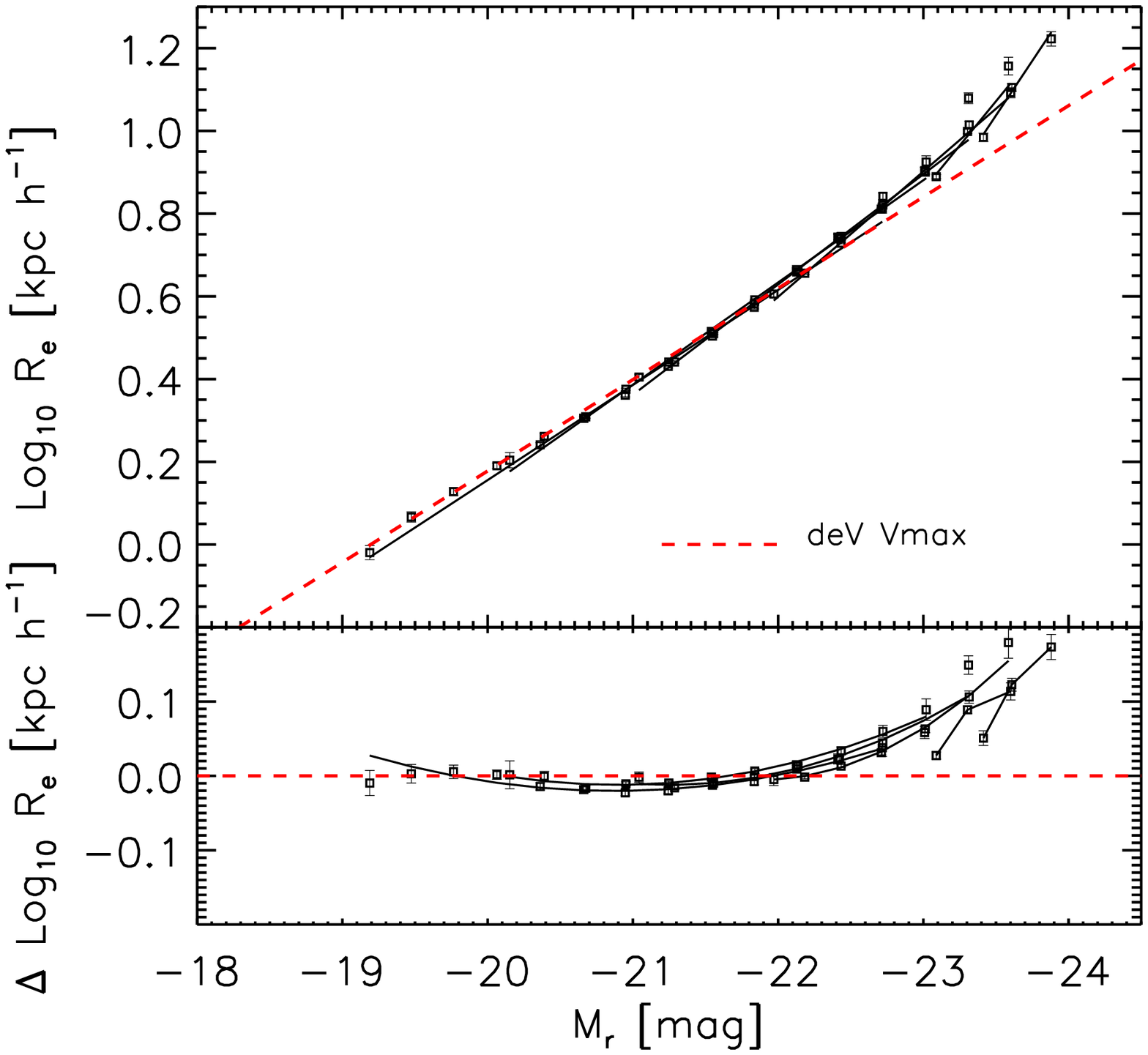}\\
 \vspace{-2cm}
 \includegraphics[width=0.95\hsize]{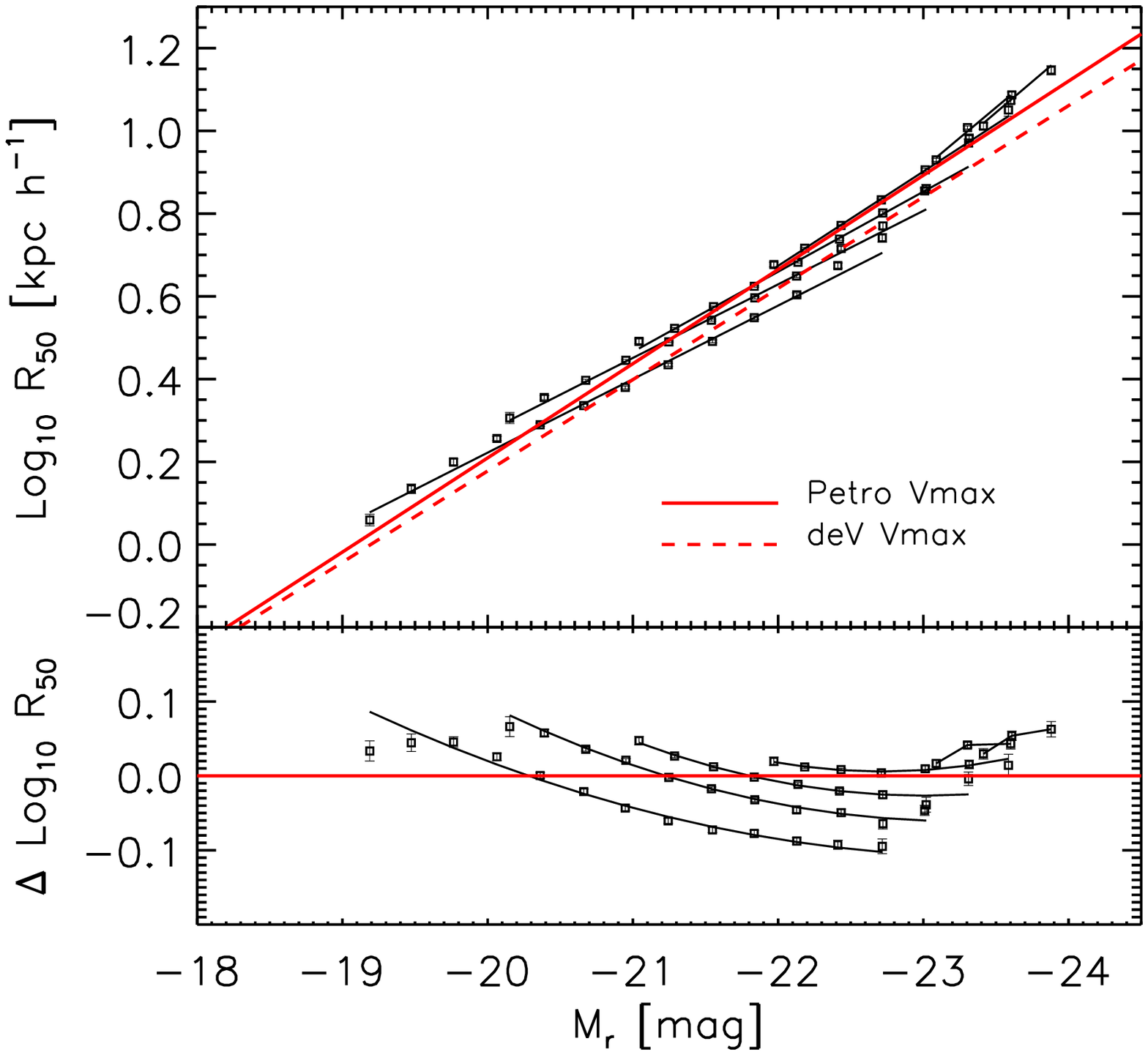}
 \caption{Residuals from the size-luminosity relation in the 
         SDSS $r$-band, derived from deVaucouleur (top) and 
         Petrosian (bottom) quantities; only the former are 
         corrected for the effects of seeing.  
         Different sets of symbols in each panel show this relation at 
         different redshifts; the relation is essentially independent 
         of redshift in the top panel, but the high redshift objects 
         appear to have larger sizes in the bottom panel.  }
 \label{rdevpet}
\end{figure}

The main text uses sizes and luminosities obtained from fitting 
deVaucouleur profiles to the images.  We chose these over the 
Petrosian quantities ({\tt R50} and {\tt petromag}) because these 
latter are not corrected for the effects of seeing.  To illustrate 
that seeing matters, Figure~\ref{rdevpet} shows the size-luminosity 
relation in six redshift bins ($0<z\le 0.07$, $0.07<z\le 0.1$, 
$0.1<z\le 0.13$, $0.15<z\le 0.18$, $0.22<z\le 0.25$ and 
$0.25<z\le 0.35$, although the luminosities are $k$- and 
$e$-corrected to $z=0$).  The top panel shows deVaucouleur- and 
bottom shows Petrosian-based results.  
In the top panel, the size-luminosity relations in the different 
redshift bins lie on top of one another.  In the bottom panel, 
however, the high redshift relations are offset to larger sizes, 
suggesting that the Petrosian sizes have been biased high by seeing.  

\begin{figure}
 \centering
 \includegraphics[width=0.95\hsize]{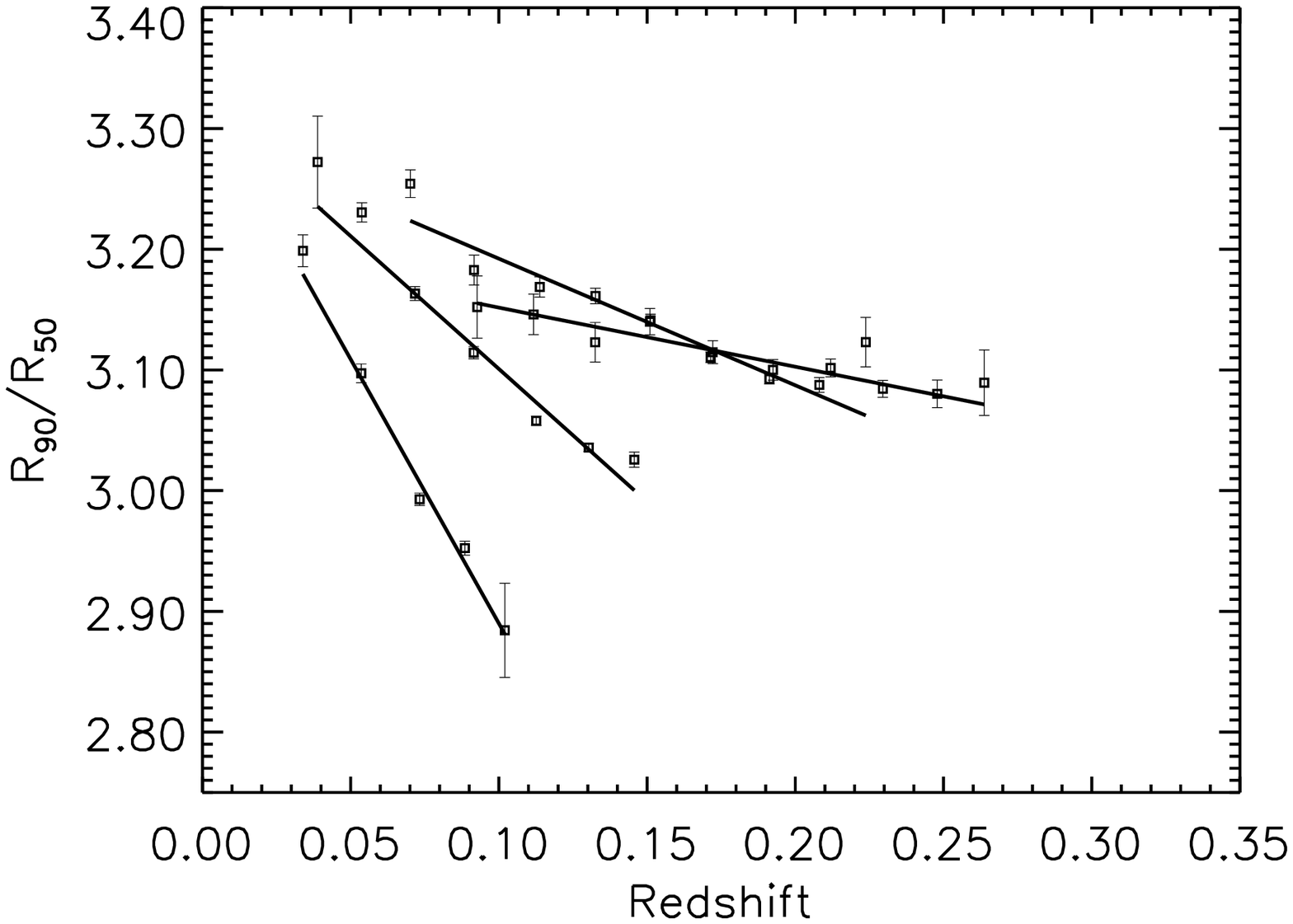}
\caption{Petrosian concentration as a function of distance from 
         observer for a few bins in (evolution corrected) luminosity  
         ($-21\le M_r\le -20.5$, $-22\le M_r\le -21.5$, $-23\le M_r\le -22.5$, and $-23.5\le M_r\le -23$)
         The higher redshift objects have smaller concentrations, 
         because {\tt R50} is increasingly affected by seeing.}
\label{r90r50}
\end{figure}

Figure~\ref{r90r50} presents other evidence that seeing matters.  
It shows the Petrosian concentration (the ratio of the radius which 
contains 90\% of the Petrosian light, to that which contains 50\%) 
as a function of redshift, for a few bins in (evolution-corrected) 
luminosity ($-21\le M_r\le -20.5$, $-22\le M_r\le -21.5$,
$-23\le M_r\le -22.5$, and $-23.5\le M_r\le -23$).  The catalog 
is apparent magnitude limited, so the most luminous bin extends to 
highest redshift.  
At fixed $L$, the higher redshift objects appear to be much less 
concentrated, but this is not a physical effect: it happens 
because {\tt R50} is increasingly affected by seeing.  

In a magnitude limited survey such as the SDSS, the more luminous 
objects are seen preferentially at larger redshifts.  If not 
accounted for, seeing effects would bias the Petrosian 
size-luminosity relation (any scaling relations for which size 
or luminosity matter would also be affected).  
This can lead to important differences in what one concludes from 
the data.  

For example, in their study of the $R-L$ relation, 
von der Linden et al. (2007) used Petrosian quantities.  
They argued that BCGs traced essentially the same $R-L$ relation 
as the bulk of the early-type galaxy population.  This 
contradicted Lauer et al. (2007) and Bernardi et al. (2007) who 
found, on the basis of seeing-corrected fits, that the BCG relation 
was substantially steeper, and used this to draw important 
conclusions about BCG formation histories.  (Steeper relations for 
BCGs have since also been found by Liu et al. 2008).  

The bottom panel of Figure~\ref{rdevpet} shows that only when 
one stacks all redshift bins together (and ignores the obvious 
offset from one redshift bin to the next) does one find a 
Petrosian $R-L$ relation that is similar to that obtained from 
seeing-corrected fits; the relation in any one redshift bin is 
significantly shallower.  
However, at the highest redshifts, where only the most luminous 
galaxies contribute, i.e., in the luminosity regime which is most 
likely to be dominated by BCGs, the $R-L$ relation is indeed 
substantially steeper.  Moreover, its slope becomes similar to 
that of the (incorrectly!) stacked sample, leading to the 
conclusion that BCGs have the same slope as the bulk of the 
population.  In effect, using Petrosian quantities without 
appropriate care for the fact that they are not seeing-corrected, 
led von der Linden et al. to confusion.  
For this reason we have not used any Petrosian-based quantities 
in our analysis, and we caution against their use in general.

This is not to say that we think deVaucouleur-based fits are 
ideal.  One might legitimately ask how the curvature we 
quantify in the main text changes if we fit to the more general 
Sersic profile.
Fitting two-dimensional Sersic profiles to the SDSS photometry 
is well beyond the scope of this work.  However, in principle, 
we could have done this approximately as follows.  
Graham et al. (2005) provide a prescription for transforming 
from Petrosian to Sersic quantities.  So we could have used these 
transformations as a proxy for actual Sersic fits, and then studied 
the associated (PSersic?!) scaling relations.  In fact, this was 
done by Desroches et al. (2007).  
We did not do this here because Figure~\ref{rdevpet} suggests 
that, because the Petrosian based quantities have not been 
seeing-corrected, this would yield biased results.

\label{lastpage}

\end{document}